\documentclass[runningheads]{llncs}
\usepackage[T1]{fontenc}
\usepackage{bm}
\usepackage{amsmath,amssymb,amsfonts}
\usepackage{color}
\usepackage{ulem}
\usepackage{subfigure}
\usepackage{caption}
\usepackage{multirow}
\usepackage{booktabs}
\usepackage{graphicx}
\usepackage{threeparttable}
\newtheorem{assumption}{Assumption}
\begin{document}
\title{PAC Infinite-Time Safety Guarantees for Stochastic Systems with Unknown Probabilistic Distributions}
\titlerunning{PAC Infinite-Time Safety Guarantees for Stochastic Systems}

\author{Taoran Wu\inst{1,2} \and
C.-H. Luke Ong\inst{3} \and
Bai Xue\inst{1,2}}

\institute{KLSS, Institute of Software, Chinese Academy of Sciences \and University of Chinese Academy of Sciences, Beijing, China \\
\and
College of Computing and Data Science, Nanyang Technological University, Singapore\\
\email{\{wutr,xuebai\}@ios.ac.cn, luke.ong@ntu.edu.sg}}

\maketitle              
\begin{abstract}
This paper studies infinite-time safety certification for discrete-time stochastic systems with continuous dynamics and unknown disturbance distributions that have full support on a known compact disturbance set. We develop a probably approximately correct (PAC) data-driven framework that uses only finitely many i.i.d. disturbance samples: with confidence at least $1-\delta$, if the certification procedure accepts a candidate certificate defining a nonempty subset of a prescribed compact safe set, then the system, starting from any state in this subset, remains within the safe set for all future times with probability at least $1-\epsilon$. Our approach builds on a fundamental characterization of infinite-time safety. We first prove that a robust invariant set exists within a compact safe set if and only if there exists an initial state from which the system remains in the safe set indefinitely with positive probability. This reduces infinite-time safety certification to two tractable subproblems: robust invariant-set computation and finite-horizon reach-avoid certification. We first compute a robust invariant set within the safe set, then use i.i.d. disturbance samples to certify, with PAC guarantees, a subset of initial states for which the probability of reaching this set within a finite horizon while remaining safe is at least a prescribed threshold. To our knowledge, this is the first PAC framework for infinite-time safety under unknown disturbance distributions. Finally, illustrative examples demonstrate the method using semi-definite programming.
\end{abstract}

\section{Introduction}
\label{sec:intro}

Safety certification is a fundamental problem in the analysis and control of stochastic dynamical systems. In many safety-critical applications, such as autonomous driving, robotic systems, and cyber-physical systems, one seeks to guarantee that the safety probability, defined as the probability that the system remains within a prescribed safe operating region over a specified operational time horizon, is at least a user-defined threshold. While finite-time safety properties are often sufficient for mission-oriented tasks, numerous applications require guarantees over an infinite time horizon, ensuring that the system never leaves the safe set throughout its entire operation. Such infinite-time safety guarantees are considerably more challenging to establish, as they require reasoning about infinitely many future evolutions of the stochastic systems. 

For stochastic systems with known probabilistic models, infinite-time safety verification has been extensively studied via probabilistic model checking and barrier certificates \cite{baier2008principles,prajna2007framework,kwiatkowska2011prism}, all of which rely on an accurate characterization of the disturbance distribution or transition probabilities. In many practical applications, however, the distribution is unavailable, and only finitely many disturbance observations or simulated outcomes can be collected. This motivates data-driven certification with rigorous probabilistic guarantees from sampled data. Recent advances in statistical learning and scenario optimization have enabled finite-time safety certification for stochastic systems with unknown disturbance distributions \cite{mathiesen2024data,wu2026pac}, establishing Probably Approximately Correct (PAC) guarantees by combining finite samples with optimization-based certificate synthesis. Yet these PAC frameworks are designed primarily for finite-horizon specifications. Extending them to the infinite horizon is far from straightforward: finite-horizon PAC guarantees are horizon-dependent and control only finitely many constraints, whereas infinite-time safety is a limiting event requiring uniform control over all time. A finite dataset constrains only finitely many disturbance realizations, so standard finite-horizon arguments do not automatically yield uniform-over-horizon guarantees. Consequently, existing finite-time techniques do not naturally generalize without additional structural assumptions or uniform concentration arguments.

This paper studies infinite-time safety certification for discrete-time stochastic systems with continuous dynamics and unknown disturbance distributions that have full support on a known compact disturbance set. We develop a data-driven certification framework that relies solely on a finite collection of i.i.d. disturbance samples. The framework yields a certification procedure with the following guarantee: with confidence at least $1-\delta$ over the sampled disturbances, whenever the procedure accepts a certificate defining a nonempty subset of a prescribed compact safe set, the infinite-horizon safety probability—i.e., the probability that the system, starting from any state in this subset, remains within the safe set over an infinite horizon—is at least $1-\epsilon$. At the core of this framework is a theoretical connection between probabilistic infinite-time safety and deterministic robust invariance. Specifically, we first establish that 
a robust invariant set—namely, a set from which the system remains safe  for all future times—exists inside a compact safe set if and only if there exists at least one initial state from which the system remains in the safe set forever with positive probability. Exploiting this equivalence, we transform infinite-time safety certification into two tractable subproblems, analogous to the decomposition established for continuous-time systems in \cite{xue2020safety}. First, we compute a robust invariant set contained within the safe set. Then, using barrier certificate methods and scenario theory \cite{calafiore2006scenario}, we perform PAC finite-time reach-avoid certification to verify that trajectories reach the invariant set within a finite time horizon while remaining safe beforehand with PAC guarantees. Finally, we demonstrate our method on several illustrative examples using semi-definite programming tools.

The main contributions of this paper are summarized as follows: 
\begin{enumerate}
    \item  \textbf{PAC framework for infinite-time safety.} We develop a PAC data-driven framework for infinite-time safety certification of discrete-time stochastic systems, using only finitely many i.i.d. disturbance samples. To the best of our knowledge, this is the first PAC certification framework for infinite-time safety of stochastic systems with unknown disturbance distributions. 

    \item \textbf{Characterization of infinite-time safety.} We reveal an intrinsic equivalence between probabilistic infinite-time safety and robust invariance.

   \item \textbf{Tractable decomposition.} Exploiting this equivalence, we reduce infinite-time safety certification to two tractable subproblems: robust invariant-set computation and finite-horizon reach-avoid certification, thereby enabling data-driven certification of infinite-time safety with PAC guarantees despite unknown disturbance distributions.
\end{enumerate}

\subsection*{Related Work}

Safety certification for stochastic discrete-time systems has been studied using statistical methods based on sampled trajectories \cite{younes2002probabilistic,sen2004statistical}, abstraction-based methods for probabilistic model checking \cite{lahijanian2015formal,kwiatkowska2011prism}, and barrier-certificate methods \cite{prajna2004safety,prajna2007framework}. Statistical methods are limited by the simulated horizon, while abstraction-based methods suffer from state-space discretization and poor scalability with dimension. Barrier certificates, in contrast, can establish finite- and infinite-horizon safety guarantees without state-space discretization. We focus on barrier-certificate methods, reviewing model-based approaches for both horizons before discussing data-driven methods and the gap in infinite-horizon certification.

\noindent\textbf{Model-based certification.}  Barrier certificates, introduced for deterministic safety and reachability verification as Lyapunov-like functions \cite{prajna2004safety,prajna2007convex}, were later extended to stochastic systems, where existing work is organized by verification horizon. For infinite-horizon verification, stochastic barrier certificates yield lower bounds on safety and reach-avoid probabilities via Ville's inequality \cite{prajna2007framework,vzikelic2023learning,zhi2024unifying}, while equation-relaxation techniques enable both lower and upper bounds \cite{xue2021reach,xue2022reach,yu2023safe,xue2026sufficient,chen2025construction}; barrier-based methods also extend to probabilistic programs and $\omega$-regular specifications \cite{chakarov2013probabilistic,mciver2017new,kenyon2021supermartingales,abate2024stochastic,wang2025verifying}. For finite-horizon verification, $c$-martingale-based conditions are standard, with early foundations in \cite{kushner1967,steinhardt2012finite}, and have since been extended to temporal-logic verification and controller synthesis under robust-invariance or stopped-process assumptions \cite{jagtap2018temporal,jagtap2020formal,santoyo2021barrier}. Refined conditions further improve finite-time probability bounds \cite{zhi2024unifying,xue2024finite}, and \cite{ghanbarpour2025characterization} studied safety verification for discrete-time systems modeled by difference inclusions.

These model-based results, however, share a common assumption: both the dynamics and the disturbance distribution are fully known, so that the expected barrier evolution or the abstraction's transition probabilities can be evaluated exactly. In many practical applications---such as autonomous systems operating in uncertain environments---this assumption fails: the disturbance distribution is unavailable and only samples can be collected. This paper targets precisely this setting, assuming known dynamics but an unknown disturbance distribution.

\noindent\textbf{Data-driven, finite-horizon certification.} Motivated by this gap, a parallel line of work has developed data-driven methods for formal verification and controller synthesis in the PAC framework, particularly for systems with limited model information. Among these methods, scenario optimization \cite{calafiore2006scenario}, which replaces infinitely many uncertainty constraints with finitely many sampled constraints, has emerged as a powerful tool for safety certification with PAC guarantees: it has been combined with barrier certificates for verification and controller synthesis of deterministic systems \cite{xue2019probably,nejati2023formal,wu2025convex,rickard2025data,samari2025data}, and was later extended to stochastic systems \cite{salamati2024data,salamati2022data,salamati2022safety}. Subsequent developments further reduced the required sample complexity through the wait-and-judge technique \cite{campi2018wait,salamati2022data} and the repetitive scenario approach \cite{calafiore2016repetitive,salamati2022safety}. These  approaches for stochastic systems typically require additional prior information, such as Lipschitz constants of the system dynamics and the barrier function, to establish their PAC guarantees. In contrast, our work considers the complementary setting in which the system dynamics are assumed to be known, while the disturbance distribution is unknown. Our approach relies solely on i.i.d. disturbance samples and does not require auxiliary quantities such as Lipschitz constants.

Closest to our work are recent studies on data-driven safety certification under unknown disturbance distributions \cite{mathiesen2023inner,gracia2024data,mathiesen2024data,wu2026pac}, which focus on finite-horizon safety. In contrast, we establish PAC guarantees for infinite-time safety by bridging finite-time reach-avoid certification and infinite-time safety through robust invariant sets. To our knowledge, this is the first PAC barrier-certificate framework providing infinite-time safety guarantees under unknown disturbance distributions. Our decomposition is related to \cite{xue2020safety}, which decomposes infinite-time safety verification for continuous-time RODEs into tractable subproblems under known disturbance distributions. In contrast, we consider discrete-time stochastic systems and provide PAC guarantees without requiring prior knowledge of the disturbance law. Furthermore, under a full-support assumption, we establish an equivalence between the existence of an initial state with positive infinite-time safety probability and the existence of a robust invariant set, which provides the theoretical basis for our decomposition.

\paragraph{Organization.} Section~\ref{sec:pre} introduces the system model and the PAC infinite-time safety certification problem. Section~\ref{sec:pacitsc} establishes the theoretical foundations of infinite-time safety and presents the proposed certification framework. Section~\ref{sec:sdp} develops a semi-definite programming implementation for polynomial systems using sum-of-squares decompositions. Section~\ref{sec:ex} demonstrates the effectiveness of the proposed approaches, and Section~\ref{sec:con} concludes this paper.

\paragraph{Notation.}
We use $\mathbb{R}^n$ to denote the $n$-dimensional Euclidean space, and $\mathbb{N}$ for the set of nonnegative integers. For two sets $\mathbb{A}$ and $\mathbb{B}$, $\mathbb{A} \setminus \mathbb{B}$ denotes the set difference, i.e., the set of all elements that belong to $\mathbb{A}$ but not to $\mathbb{B}$. The notation $\overline{\mathbb{A}}$ denotes the closure of $\mathbb{A}$.  Given a set $A\subseteq\mathbb{R}^k$, $1_{A}(\cdot)$ denotes the indicator function of $A$, i.e., $1_{A}(\bm{y}):=1$ if $\bm{y}\in A$ and $0$ otherwise. 
Given a probability space $(\Omega, \mathcal{F}, \textnormal{P}_{\bm{y}})$, we denote by $\textnormal{P}_{\bm{y}}[\mathbb{E}]$ the probability of an event $\mathbb{E} \in \mathcal{F}$, and by $\textnormal{E}_{\bm{y}}[\cdot]$ the expectation with respect to $\textnormal{P}_{\bm{y}}$. $\mathbb{R}[\bm{y}]$ denotes the ring of polynomials in variables $\bm{y}$ and $\Sigma[\bm{y}]$ denotes the set of sum-of-squares (SOS) polynomials, i.e., $\Sigma[\bm{y}]=\left\{p\in\mathbb{R}[\bm{y}]\;\middle|\;\exists k\in\mathbb{N}, \exists q_1,\ldots,q_k\in \mathbb{R}[\bm{y}], \text{such that}~ p(\bm{y})=\sum_{i=1}^{k}q_i^2(\bm{y})\right\}$.
\section{Preliminaries}
\label{sec:pre}
In this section, we introduce stochastic systems and the PAC infinite-time safety certification problem of interest.

\paragraph{System model.}
We consider a discrete-time stochastic system
\begin{equation}
    \bm{x}(j+1) = \bm{f}(\bm{x}(j), \bm{d}(j)), \quad j \in \mathbb{N},
    \label{eq:system}
\end{equation}
where $\bm{x}(j) \in \mathbb{R}^n$ is the system state and $\bm{d}(j) \in \mathbb{D}$ is a stochastic disturbance with $\mathbb{D}$ being compact in $\mathbb{R}^{d}$. The update map $\bm{f}(\cdot,\cdot): \mathbb{R}^n \times \mathbb{D} \to \mathbb{R}^n$ is assumed known,  deterministic, and continuous over $(\bm{x},\bm{d})$. The disturbance sequence $\{\bm{d}(j)\}_{j\in \mathbb{N}}$ consists of i.i.d. random vectors on a probability space $(\mathbb{D},\mathcal{F},\text{P}_{\bm{d}})$, with $\bm{d}(j)\sim \text{P}_{\bm{d}}$ for all $j\in \mathbb{N}$. The expectation $\textnormal{E}_{\bm{d}}[\cdot]$ is taken with respect to $\text{P}_{\bm{d}}$. Further, we assume that the support of $\mathrm{P}_{\bm{d}}$ is exactly $\mathbb{D}$, i.e., for every $\bm{d}\in\mathbb{D}$ and every $r>0$, $\mathrm{P}_{\bm{d}}[\mathbb{B}_r(\bm{d})]>0$, where $\mathbb{B}_r(\bm{d})$ denotes the open ball of radius $r$. 

\begin{definition}[Disturbance Signal]  
A disturbance signal is a sequence $\pi=\{\bm{d}(j)\}_{j \in \mathbb{N}}$, i.e., a sample path (realization) of the stochastic disturbance process, where $\bm{d}(j)$ denotes the $j$-th coordinate of $\pi$. The canonical sequence space $\mathbb{D}^{\infty}$ is equipped with the product $\sigma$-algebra $\mathcal{B}(\mathbb{D}^{\infty})=\bigotimes_{j=0}^{\infty} \mathcal{B}(\mathbb{D})$, which coincides with the Borel $\sigma$-algebra generated by the product topology since $\mathbb{D}$ is a compact metric space, and with the product probability measure, which we denote by $\textnormal{P}_{\pi}:=\textnormal{P}_{\bm{d}}^{\infty}$.   
\end{definition}  

\begin{definition}[Trajectory]  
Given an initial state $\bm{x}_0 \in \mathbb{R}^n$ and $\pi=\{\bm{d}(j)\}_{j\in\mathbb{N}}\in\mathbb{D}^{\infty}$, the corresponding trajectory of system \eqref{eq:system} is the mapping $\bm{\phi}_{\pi}^{\bm{x}_0}(\cdot) \colon \mathbb{N} \rightarrow \mathbb{R}^n$ defined recursively as  
\[
\begin{cases}
    \bm{\phi}_{\pi}^{\bm{x}_0}(j+1) = \bm{f}(\bm{\phi}_{\pi}^{\bm{x}_0}(j), \bm{d}(j)), \quad \forall j\in \mathbb{N},\\
    \bm{\phi}_{\pi}^{\bm{x}_0}(0) = \bm{x}_0.
\end{cases}
\]  
\end{definition}

\paragraph{$\infty$-step safety certification.}
A compact set $\mathbb{X} \subseteq \mathbb{R}^n$ defines the safe region. Given an initial state $\bm{x} \in \mathbb{X}$, the system is $\infty$-step safe with respect to a threshold $\epsilon\in [0,1)$ if the infinite-horizon safety probability is no less than $1-\epsilon$, i.e.,
\[
    \mathrm{P}_{\pi}\Bigl[\, \bigwedge_{j \in \mathbb{N}} \bm{\phi}^{\bm{x}}_{\pi}(j) \in \mathbb{X} \,\Bigr] \geq 1-\epsilon.
\]

In this paper, rather than assuming knowledge of the disturbance law $\mathrm{P}_{\bm{d}}$, we assume access to a sampling mechanism that returns independent draws from it. Concretely, we are given a finite collection
\[
    D^M = \bigl\{ \bm{d}^{(1)}, \dots, \bm{d}^{(M)} \bigr\},
    \qquad \bm{d}^{(i)} \overset{\text{i.i.d.}}{\sim} \mathrm{P}_{\bm{d}},
\]
where the notation $\bm{d}^{(i)} \overset{\text{i.i.d.}}{\sim} \mathrm{P}_{\bm{d}}$ specifies the law governing the samples, not a quantity available to the practitioner: the draws are observed, whereas $\mathrm{P}_{\bm{d}}$ remains unknown. We write $\mathrm{P}_{\bm{d}}^{M}$ for the $M$-fold product distribution of $\mathrm{P}_{\bm{d}}$ on $\mathbb{D}^{M}$, so that $D^{M} \sim \mathrm{P}_{\bm{d}}^{M}$.

\begin{problem}[PAC Infinite-Time Safety Certification]
\label{prob:1}
Given a confidence parameter $\delta\in (0,1)$ and a probability threshold $\epsilon\in [0,1)$, the goal is to design a certification procedure $\mathcal{A}$ that provides $\infty$-step safety certification in the PAC sense: with confidence at least $1-\delta$ over the sample set $D^M$, if $\mathcal{A}(D^M)$ accepts ---thereby \ returning a certificate that defines a nonempty subset $\mathbb{X}_s(D^M) \subseteq \mathbb{X}$ of the safe region---the system \eqref{eq:system}, starting from $\bm{x} \in \mathbb{X}_s$, remains in $\mathbb{X}$ for all future times with probability at least $1-\epsilon$. 

Formally, the $\infty$-step PAC guarantee can be expressed as
\[
\textnormal{P}_{\bm{d}}^M \Bigg[
\mathcal{A}(D^M) \Rightarrow \left(\forall \bm{x}\in \mathbb{X}_s(D^M),
\textnormal{P}_{\pi}\Big[\bigwedge_{j=1}^{\infty} \bm{\phi}^{\bm{x}}_{\pi}(j) \in \mathbb{X} \Big] \ge 1-\epsilon
\right)
\Bigg] \ge 1-\delta, 
\]
\end{problem}

In the sequel, we denote by $\widehat{\mathbb{X}}$ a compact set containing $\mathbb{X}$ and all states reachable in one step from  $\mathbb{X}$ by system \eqref{eq:system}, i.e., 
    \[\widehat{\mathbb{X}} \supseteq \{\bm{y}\mid \bm{y}=\bm{f}(\bm{x},\bm{d}), \bm{x}\in \mathbb{X},\bm{d}\in \mathbb{D}\} \cup \mathbb{X}.\]
  Such a set always exists: since $\bm{f}(\bm{x},\bm{d})$ is continuous over $(\bm{x},\bm{d})$ and $\mathbb{X}\times \mathbb{D}$ is compact, the set on the right-hand side is compact.

\section{PAC Infinite-Time Safety Certification}
\label{sec:pacitsc}
In this section, we propose a data-driven framework for Problem~\ref{prob:1}. By establishing an equivalence between infinite-time safety and the existence of robust invariant sets, we decompose the problem into two tractable subproblems: robust invariant-set computation and PAC finite-time reach-avoid certification, both within a unified barrier-certificate framework. Specifically, we first compute a robust invariant set within the safe set and then use finitely many i.i.d. disturbance samples to certify that the probability of reaching this set within a finite horizon while remaining safe exceeds a prescribed threshold. The latter step combines robust barrier certificates with scenario optimization.

\subsection{From Infinite-Time Safety to Invariant Sets}
We first establish a structural characterization of infinite-time safety in terms of robust invariant sets. This result provides the theoretical basis for reducing infinite-time safety certification to robust invariant-set computation and finite-time reach-avoid certification.

The intuition comes from the disturbance-free counterpart of \eqref{eq:system}, i.e., $\bm{x}(j+1)=\bm{f}(\bm{x}(j))$. If an initial state keeps the system within the compact safe set $\mathbb{X}$ for all future times, then its trajectory forms an invariant subset of $\mathbb{X}$, which the system never leaves when initialized at any of its states. The converse also holds. For stochastic systems, the analogous notion is almost-sure probabilistic invariance. In fact, for weakly continuous Markov processes over compact sets, a nonempty almost-sure probabilistic invariant subset exists if and only if some initial state has a positive probability of remaining in the safe set indefinitely \cite{tkachev2014characterization}. Since our system induces a weakly continuous transition kernel, this characterization applies. For completeness, we state the result below and provide an alternative proof tailored to our system model.

\begin{lemma}[Equivalence Between Almost-Sure Invariance and Positive Safety Probability]
\label{thm:aps}
The following statements are equivalent.
\begin{enumerate}
    \item[(1)] There exists a nonempty $\mathbb{S}\subseteq \mathbb{X}$ such that $\mathrm{P}_{\pi}\Big[ \bigwedge_{j=0}^{\infty} \bm{\phi}^{\bm{x}}_{\pi}(j) \in \mathbb{S}\Big]=1, \forall \bm{x}\in \mathbb{S}$.
    \item[(2)] There exists a state $\bm{x}\in \mathbb{X}$ satisfying
    $\mathrm{P}_{\pi}\Big[\bigwedge_{j=1}^{\infty} \bm{\phi}^{\bm{x}}_{\pi}(j)\in \mathbb{X}\Big]>0$.
\end{enumerate}
\end{lemma}
\begin{proof}
The proof is shown in Appendix \ref{sec:pro}. \qed
\end{proof}

Based on Lemma \ref{thm:aps}, we show that, under a mild regularity assumption on the disturbance distribution, every such set is in fact robustly invariant. This link is essential for what follows, since robust invariant sets can be computed from the dynamics and the disturbance support alone, without knowledge of the disturbance distribution.

\begin{theorem}[Equivalence Between Robust Invariance and Positive Safety Probability]
\label{thm:ris}
The following statements are equivalent.
\begin{enumerate}
    \item[(1)] There exists a nonempty compact robust invariant set $\mathbb{S}\subseteq\mathbb{X}$, i.e.,
    \[\bm{f}(\bm{x},\bm{d})\in\mathbb{S},\qquad \forall(\bm{x},\bm{d})\in\mathbb{S}\times\mathbb{D},\]
    or equivalently,
    $\bm{\phi}_{\pi}^{\bm{x}}(k)\in\mathbb{S}, \forall(\bm{x},\pi,k)\in\mathbb{S}\times\mathbb{D}^{\infty}\times\mathbb{N}$.
    \item[(2)] There exists a state $\bm{x}\in\mathbb{X}$ satisfying
    $\mathrm{P}_{\pi}\Big[\bigwedge_{k=0}^{\infty}\bm{\phi}_{\pi}^{\bm{x}}(k)\in\mathbb{X}\Big]>0$.
\end{enumerate}
\end{theorem}
\begin{proof}
The proof is shown in Appendix \ref{sec:pro}. \qed
\end{proof}

Theorem~\ref{thm:ris} establishes the theoretical foundation of our framework: the infinite-horizon safety probability is positive for some initial state if and only if the prescribed safe set contains a nonempty robust invariant set. This equivalence decomposes infinite-time safety certification into two tasks: (i) computing a robust invariant set contained in the safe set, and (ii) finite-horizon PAC reach-avoid certification, i.e., certifying with PAC guarantees that trajectories reach this set within a finite horizon while remaining safe beforehand. The maximal robust invariant set---the union of all robust invariant sets---can be characterized by a Bellman-type equation \cite{xue2021robust}, but such equations are generally intractable. We therefore compute a robust invariant set by solving an optimization problem over a barrier certificate $v:\mathbb{R}^n\to\mathbb{R}$ satisfying
\begin{equation}
\label{robust}
    \begin{cases}
        v(\bm{f}(\bm{x},\bm{d})) \geq \gamma\, v(\bm{x}), &\forall (\bm{x},\bm{d}) \in \mathbb{X}\times \mathbb{D},\\[2pt]
        v(\bm{x})<0, &\forall \bm{x}\in \widehat{\mathbb{X}} \setminus \mathbb{X},
    \end{cases}
\end{equation}
whose nonnegative superlevel set $\mathbb{S}=\{\bm{x}\in \widehat{\mathbb{X}} \mid v(\bm{x})\geq 0\}$ is a robust invariant subset of the prescribed safe set, where $\gamma \in (0,1]$ is a user-specified scalar. To search for such a function, we consider a parameterized certificate $v(\bm{a},\cdot):\mathbb{R}^n\to\mathbb{R}$ and seek a parameter vector $\bm{a}=(\bm{a}[1],\ldots,\bm{a}[m_r])^{\top}\in\mathbb{R}^{m_r}$ such that $v(\bm{a},\bm{x})$ satisfies \eqref{robust}. This search can be reformulated as the following uncertain optimization problem:
\begin{equation}
\begin{split}
    &\textstyle\min_{\bm{a}\in \mathbb{R}^{m_r}} \; -\frac{1}{N_r}\sum_{i=1}^{N_r} v(\bm{a},\bm{x}'_i)\label{eq:robust_invariant_set} \\
    \text{s.t.}\;&
    \begin{cases}
        v(\bm{a},\bm{f}(\bm{x},\bm{d})) \geq \gamma\, v(\bm{a},\bm{x}), &\forall (\bm{x},\bm{d}) \in \mathbb{X}\times \mathbb{D},\\[2pt]
        v(\bm{a},\bm{x})<0, &\forall \bm{x}\in \widehat{\mathbb{X}} \setminus \mathbb{X},
    \end{cases}
\end{split}
\end{equation}
where $\{\bm{x}'_i\}_{i=1}^{N_r}$ is a family of states distributed approximately uniformly over the safe set $\mathbb{X}$, which can be generated by random sampling from the uniform distribution on $\mathbb{X}$.

\begin{remark}
   The ideal objective of the program \eqref{eq:robust_invariant_set} is to find a function $v(\bm{a}, \bm{x})$ that is maximized over $\mathbb{X}$, 
   thereby encouraging a large set $\mathbb{S}$.
   A natural choice for the objective function is to maximize $\int_{\mathbb{X}} v(\bm{a}, \bm{x}) d\bm{x}$. However, since this integral may not be analytically tractable, we approximate it using numerical sampling $\frac{1}{N_r}\sum_{i=1}^{N_r}v(\bm{a},\bm{x}'_i)$. Importantly, the states $\{\bm{x}'_i\}_{i=1}^{N_r}$ only facilitate this approximation: they are drawn and then fixed, and they play no role in the PAC characterization, which depends solely on the disturbance samples.
\end{remark}

In the remainder of this paper, we focus primarily on the second task, developing PAC data-driven reach-avoid certification.

\subsection{Finite-Time PAC Reach-Avoid Certification}
\label{sec:ftrac}
In this subsection, we develop PAC data-driven approaches for finite-time reach-avoid certification, which, together with the robust invariant-set computation of the previous subsection, provide a complete solution to infinite-time safety certification. Our framework integrates scenario optimization with robust barrier-like functions, termed robust barrier-guidance functions (RBGFs), for finite-time reach-avoid certification. We first review RBGFs for discrete-time systems.

\begin{definition}[Robust Barrier-Guidance Function]
\label{def:rbgf}
Given $\gamma\in(1,\infty)$, a safe set $\mathbb{X}$, and a target set $\mathbb{S}\subseteq \mathbb{X}$, consider a function $h:\mathbb{R}^n\to\mathbb{R}$ that is upper-bounded on $\mathbb{X}$, and let $\mathbb{X}_s:=\{\bm{x}\in\mathbb{X}\mid h(\bm{x})>0\}$. Then $h$ is called a robust barrier-guidance function (RBGF) for system~\eqref{eq:system} if $\mathbb{X}_s\neq\emptyset$ and
\begin{equation}
\label{eq:rbf}
\begin{cases}
    h(\bm{x})\le 0, & \forall \bm{x}\in\overline{\widehat{\mathbb{X}}\setminus\mathbb{X}},\\[2pt]
    h(\bm{f}(\bm{x},\bm{d}))\ge \gamma\, h(\bm{x}), & \forall(\bm{x},\bm{d})\in(\mathbb{X}\setminus\mathbb{S})\times\mathbb{D}.
\end{cases}
\end{equation}
\end{definition}

\begin{proposition}[\cite{zhao2022inner}]
\label{prop:rbgf}
Let $h:\mathbb{R}^n\to\mathbb{R}$ be an RBGF for system~\eqref{eq:system} with parameter $\gamma>1$. Define $\mathbb{X}_s:=\{\bm{x}\in\mathbb{X}\mid h(\bm{x})>0\}$ and $\mathbb{X}_{s,\beta}:=\{\bm{x}\in\mathbb{X}\mid h(\bm{x})\ge\beta\}$. Then, for every disturbance signal $\pi\in\mathbb{D}^{\infty}$ and every initial state $\bm{x}_0\in\mathbb{X}_s$, the trajectory of system~\eqref{eq:system} reaches $\mathbb{S}$ in finite time while remaining in $\mathbb{X}_s$ prior to reaching $\mathbb{S}$. Moreover, if $\sup_{\bm{x}\in\mathbb{X}\setminus \mathbb{S}}h(\bm{x})\le H$ and $\beta\in(0,H)$, then for every $\pi\in\mathbb{D}^{\infty}$ and every $\bm{x}_0\in\mathbb{X}_{s,\beta}$, the trajectory reaches $\mathbb{S}$ within at most $\lfloor\log_{\gamma}(H/\beta)\rfloor+1$ steps while remaining in $\mathbb{X}_s$ prior to reaching $\mathbb{S}$.
\end{proposition}

To search for such a function, we consider a parameterized function $h(\bm{a}, \cdot): \mathbb{R}^n \rightarrow \mathbb{R}$ with $h(\bm{0},\bm{x})\equiv 0$ and  find the parameter vector $\bm{a} = (\bm{a}[1], \ldots, \bm{a}[m])^{\top} \in \mathbb{R}^m$ such that $h(\bm{a}, \bm{x})$ satisfies the constraints \eqref{eq:rbf}. This search problem can be reformulated as the following uncertain optimization problem:
\begin{equation}
\begin{split}
    &\textstyle\min_{\bm{a}\in \mathbb{R}^m } \quad - \frac{1}{N_o}\textstyle\sum_{i=1}^{N_o}h(\bm{a},\bm{x}'_i)\label{eq:RBF_linear1} \\
    \text{s.t.}&\begin{cases}
        h(\bm{a},\bm{x})\leq 0, &\forall \bm{x}\in \overline{\widehat{\mathbb{X}}\setminus \mathbb{X}},\\
        h(\bm{a}, \bm{f}(\bm{x},\bm{d})) \geq \gamma h(\bm{a}, \bm{x}), &\forall \bm{x} \in \mathbb{X}\setminus \mathbb{S}, \forall \bm{d} \in \mathbb{D},\\
        \bm{a}\in[-U_a,U_a]^{m}, 
    \end{cases} 
\end{split}
\end{equation}
where $\gamma \in (1,\infty)$ and $U_a > 0$ are user-defined scalar values, and $\{\bm{x}'_i\}_{i=1}^{N_o}$ is a family of specified states distributed evenly over $\mathbb{X}\setminus \mathbb{S}$, which can be generated by random sampling from a uniform distribution over the set $\mathbb{X}\setminus \mathbb{S}$. The states $\{\bm{x}'_i\}_{i=1}^{N_o}$ play the same role as in \eqref{eq:robust_invariant_set}: they approximate the intractable integral $\int_{\mathbb{X}\setminus\mathbb{S}}h(\bm{a},\bm{x})\,\mathrm{d}\bm{x}$ by the empirical average in the objective, steering the optimization toward a large $\mathbb{X}_s$. They are drawn once and then fixed, and play no role in the PAC characterization, which depends solely on the disturbance samples.

The optimization problem \eqref{eq:RBF_linear1} always admits a feasible solution, since $\bm{a}\equiv\bm{0}$ yields $h(\bm{0},\bm{x})\equiv 0$ and satisfies all constraints. Let $\bm{a}^*$ denote a feasible solution. If $\mathbb{X}_s=\{\bm{x}\in\mathbb{X}\mid h(\bm{a}^*,\bm{x})>0\}\neq\emptyset$, then $h(\bm{a}^*,\bm{x})$ is an RBGF. However, problem \eqref{eq:RBF_linear1} enforces worst-case satisfaction over all possible disturbances, which can render its nontrivial solutions overly conservative or may preclude the existence of any solution with $\mathbb{X}_s\neq\emptyset$. To alleviate this issue, we adopt a probabilistic formulation of the safety certification problem, which provides quantitative safety guarantees instead of relying on excessively conservative worst-case bounds. Since the true disturbance distribution is unknown, such probabilistic characterizations cannot be computed directly. We therefore resort to a data-driven approach, leveraging finitely many disturbance samples to obtain confidence-guaranteed probabilistic safety assurances. Specifically, we draw $M$ i.i.d.\ disturbance samples $D^M=\{\bm{d}^{(j)}\}_{j=1}^{M}$, with each $\bm{d}^{(j)}$ drawn independently from $\mathrm{P}_{\bm{d}}$, and replace the universal disturbance constraints in \eqref{eq:RBF_linear1} by their sample-based counterparts, leading to the following sample-based optimization problem:
\begin{equation}
\begin{split}
    &\textstyle\min_{\bm{a}\in \mathbb{R}^m} \quad -\frac{1}{N_o}\sum_{i=1}^{N_o}h(\bm{a},\bm{x}'_i)\label{eq:RBF_linear21} \\
    \text{s.t.}\;&
    \begin{cases}
        h(\bm{a},\bm{x})\leq 0, &\forall \bm{x}\in \overline{\widehat{\mathbb{X}}\setminus \mathbb{X}},\\[2pt]
        h(\bm{a},\bm{f}(\bm{x},\bm{d}^{(j)}))\geq \gamma\, h(\bm{a},\bm{x}), &\forall \bm{x}\in\mathbb{X}\setminus\mathbb{S},\ j=1,\ldots,M,\\[2pt]
        \bm{a}\in[-U_a,U_a]^{m}.
    \end{cases}
\end{split}
\end{equation}

Similar to program \eqref{eq:RBF_linear1}, program \eqref{eq:RBF_linear21} is always feasible. To obtain PAC finite-time reach-avoid guarantees, we employ the scenario approach \cite{calafiore2006scenario}, which requires the constraints to be convex in the decision variable.

\begin{assumption}
\label{as:h}
The parameterized function $h(\bm{a},\cdot):\mathbb{R}^n\to\mathbb{R}$ is linear in the parameter $\bm{a}\in\mathbb{R}^m$ and continuous in $\bm{x}\in\mathbb{R}^n$; equivalently, there exists a continuous feature map $\bm{\phi}:\mathbb{R}^n\to\mathbb{R}^m$ such that
\[
h(\bm{a},\bm{x})=\bm{a}^{\top}\bm{\phi}(\bm{x}),\qquad \forall(\bm{a},\bm{x})\in\mathbb{R}^m\times\mathbb{R}^n.
\]
\end{assumption}

Under Assumption~\ref{as:h}, the constraints of \eqref{eq:RBF_linear21} are affine in $\bm{a}$, so the scenario approach applies. The following lemma uses it to bound the one-step barrier increment of the synthesized candidate RBGF, uniformly over the state space, from finitely many disturbance samples.

\begin{lemma}[State-uniform PAC one-step increment guarantee]
\label{lemma:robust_s}
    Let $\epsilon, \delta \in (0,1)$ and $\gamma \in (1,\infty)$ be given constants. 
    Let $\bm{a}^*(D^M)$ be the optimal solution to \eqref{eq:RBF_linear21}, selected according to the tie-break rule as suggested in  \cite{calafiore2006scenario} when the program \eqref{eq:RBF_linear21} admits multiple minimizers, 
    and define the certification procedure
\begin{equation}
\label{cp}
\mathcal{A}(D^M) := 
\begin{cases}
1, & \text{if } ~{\mathbb{X}_{s,\beta}\neq \emptyset},\\
0, & \text{otherwise.}
\end{cases}
\end{equation} Then, under Assumption \ref{as:h}, the probabilistic guarantee holds for every $\beta>0$:
  \begin{equation*}
\begin{split}
   &\textnormal{P}_{\bm{d}}^M \left[
    \begin{split}
     {\mathcal{A}(D^M)} \Rightarrow \left(\forall \bm{x}\in \mathbb{X}_{s,\beta}\setminus \mathbb{S}, \textnormal{P}_{\bm{d}}\left[
     \begin{split}
     &h(\bm{a}^*(D^M),\bm{f}(\bm{x},\bm{d})) \geq \\
     &\gamma h(\bm{a}^*(D^M),\bm{x})  \bigwedge  \bm{f}(\bm{x},\bm{d}) \in \mathbb{X}_s
     \end{split}
     \right]\geq 1-\epsilon \right)
        \end{split}
        \right]\\
        &\geq 
          1-\delta,
\end{split}
\end{equation*}
where $\mathbb{X}_{s,\beta}=\{\bm{x}\in \mathbb{X}\mid h(\bm{a}^*(D^M),\bm{x}) \geq \beta\}$ and the number $M$ satisfies \begin{equation}
    \label{MM_so}
    M \geq \frac{2}{\epsilon}(\ln{\frac{1}{\delta}}+m).
\end{equation}
\end{lemma}
\begin{proof}
 The proof follows that of Lemma~2 in \cite{wu2026pac}. \qed
\end{proof}

The one-step guarantee of Lemma~\ref{lemma:robust_s} is the building block for the finite-horizon result. We now lift it to a finite-time reach-avoid guarantee by applying it recursively along the trajectory and invoking the finite-time reachability property of RBGFs.

\begin{lemma}[State-uniform PAC Finite-time Reach-avoid Guarantee]
\label{lemma:so}
    Let $\epsilon, \delta  \in (0,1)$ and $\gamma \in (1,\infty)$ be given constants.
Let $\bm{a}^*(D^{M})$ denote the optimal solution to \eqref{eq:RBF_linear21}, selected according to the tie-break rule as suggested in  \cite{calafiore2006scenario} when the program \eqref{eq:RBF_linear21} admits multiple minimizers. Define the certification procedure $\mathcal{A}(D^M)$ as in \eqref{cp}. Then, under Assumption \ref{as:h}, the probabilistic guarantee holds for every $\beta>0$:
 \begin{equation*}
\begin{split}
   &\textnormal{P}_{\bm{d}}^M \left[
    {\mathbb{X}_{s,\beta}\neq \emptyset} \Rightarrow
    \begin{split}
         \left(\forall \bm{x}\in \mathbb{X}_{s,\beta}\setminus \mathbb{S}, \textnormal{P}_{\pi}\left[
         \begin{split}
         &\exists k \leq K. \bm{\phi}_{\pi}^{\bm{x}} (k)\in \mathbb{S}~\wedge\\
         &\bigwedge_{j=0}^{k-1}\bm{\phi}_{\pi}^{\bm{x}}(j) \in \mathbb{X}_{s,\beta}
         \end{split}
         \right]\geq (1-\epsilon)^{K}\right)
        \end{split}
        \right]\\
        &\geq 
          1-\delta,
\end{split}
\end{equation*}
where $\mathbb{X}_{s,\beta}=\{\bm{x}\in \mathbb{X}\mid h(\bm{a}^*(D^M),\bm{x}) \geq \beta\}$,  $\sup_{\bm{x}\in \mathbb{X}\setminus\mathbb{S}} h(\bm{a}^*(D^M),\bm{x}) \leq H$, $K=\lfloor{\log_{\gamma}\frac{H}{\beta}\rfloor}+1$, and the sample size $M$ satisfies \eqref{MM_so}.
\end{lemma}
\begin{proof}
The proof is shown in Appendix \ref{sec:pro}.   
    \qed
\end{proof}

\begin{remark}[Uniform versus state- and trajectory-dependent guarantees]
\label{rem:sharpen}
The bound of Lemma~\ref{lemma:so} is uniform over $\bm{x}\in\mathbb{X}_{s,\beta}$ and therefore reflects the worst case, attained at states with $h(\bm{a}^*(D^M),\bm{x})=\beta$.  However, at the level of states, for any $\beta'\ge\beta$, applying Lemma~\ref{lemma:so} with threshold $\beta'$ certifies the smaller set $\mathbb{X}_{s,\beta'}$ with horizon $K'=\lfloor\log_{\gamma}(H/\beta')\rfloor+1\le K$ and probability $(1-\epsilon)^{K'}\ge(1-\epsilon)^{K}$.

The same monotonicity governs the evolution along a trajectory, where the remaining guarantee is determined by the realized barrier value rather than by the elapsed time. Let $\mathcal{E}_k$ denote the event that the growth condition holds at each of the first $k$ steps, so that $\mathrm{P}_\pi[\mathcal{E}_k]\ge(1-\epsilon)^k$. On $\mathcal{E}_k$, unrolling the growth constraint gives $h(\bm{a}^*(D^M),\bm{\phi}^{\bm{x}}_{\pi}(k))\ge\gamma^{k}h(\bm{a}^*(D^M),\bm{x})$, so the remaining horizon from the current state $K_k=\big\lfloor\log_{\gamma}\big(H/h(\bm{a}^*(D^M),\bm{\phi}^{\bm{x}}_{\pi}(k))\big)\big\rfloor+1$ 
satisfies $K_k\le K_0-k$, where $K_0=\lfloor\log_{\gamma}(H/h(\bm{a}^*(D^M),\bm{x}))\rfloor+1$. Hence the remaining reach probability is at least $(1-\epsilon)^{K_k}$, which is no less than $(1-\epsilon)^{K_0-k}$. \qed
\end{remark}

The finite-time reach-avoid guarantee, together with the invariance of the target set, yields an infinite-time safety guarantee. Indeed, once the trajectory reaches the robust invariant set $\mathbb{S}$, Theorem \ref{thm:ris} ensures that it remains there, and hence in the safe set, for all future times. Combining the two ingredients gives the following PAC infinite-time safety guarantee.

\begin{theorem}[PAC Infinite-Time Safety Guarantee]
\label{thm:so}
Let $\epsilon,\delta\in(0,1)$ and $\gamma\in(1,\infty)$ be given constants, and let $M$ satisfy \eqref{MM_so}. Let $\bm{a}_{h}^*(D^{M})$ be an optimal solution to \eqref{eq:RBF_linear21}, selected according to the tie-break rule as suggested in  \cite{calafiore2006scenario} when the program \eqref{eq:RBF_linear21} admits multiple minimizers, and 
\[
\mathbb{S}=\big\{\bm{x}\in\widehat{\mathbb{X}}\mid v(\bm{a}_{v}^*,\bm{x})\ge0\big\}\subseteq\mathbb{X}
\]
be the robust invariant set, which is independent of $D^M$ and computed via solving \eqref{eq:robust_invariant_set}. Assume that $\mathbb{S}\neq\emptyset$. Define 
$\mathbb{X}_{s,\beta}(D^{M})=\big\{\bm{x}\in\mathbb{X}\mid h(\bm{a}_{h}^*(D^{M}),\bm{x})\ge\beta\big\}$ with $\beta > 0$.
Then, under Assumption~\ref{as:h}, the following hold:
\begin{enumerate}
    \item[(i)] For every $\bm{x}\in\mathbb{S}$, $ \mathrm{P}_{\pi}\big[\bm{\phi}_{\pi}^{\bm{x}}(j)\in\mathbb{X}\ \text{for all }j\in\mathbb{N}\big]=1$.
    \item[(ii)] With probability at least $1-\delta$ over $D^{M}$, if the certification procedure accepts, i.e., if $\mathbb{X}_{s,\beta}(D^{M})\neq\emptyset$, then
    \begin{equation}
    \label{eq:pac_inf}
    \forall\bm{x}\in\mathbb{X}_{s,\beta}(D^{M})\setminus\mathbb{S}:\quad
    \mathrm{P}_{\pi}\big[\bm{\phi}_{\pi}^{\bm{x}}(j)\in\mathbb{X}\ \text{for all }j\in\mathbb{N}\big]\ge(1-\epsilon)^K,
    \end{equation}
    where $H=\sup_{\bm{x}\in\mathbb{X}\setminus\mathbb{S}}h(\bm{a}_{h}^*(D^{M}),\bm{x})$ and $K=\Big\lfloor\log_{\gamma}\frac{H}{\beta}\Big\rfloor+1$.
\end{enumerate}
\end{theorem}

The above theorem provides a PAC infinite-time safety guarantee for system~\eqref{eq:system} that requires no knowledge of the disturbance distribution. 

\section{Semi-definite Programming Implementation}
\label{sec:sdp}
In this section, we develop a semi-definite programming procedure for certifying infinite-horizon safety of polynomial systems satisfying Assumption \ref{assump2}.

\begin{assumption}
\label{assump2}
\begin{enumerate}
    \item The safe set $\mathbb{X}$, the one-step inclusion set $\widehat{\mathbb{X}}$, and the disturbance set $\mathbb{D}$ are semi-algebraic sets of the following form:
\[
\mathbb{X}=\{\bm{x}\in\mathbb{R}^n\mid g(\bm{x})\leq 0\},~\widehat{\mathbb{X}} =
\{\bm{x}\in\mathbb{R}^n\mid \hat g(\bm{x})\leq 0\}, ~\mathbb{D}
=\{\bm{d}\in\mathbb{R}^d\mid p(\bm{d})\leq 0\},
\]
where $g$, $\hat g \in \mathbb{R}[\bm{x}]$, and $p\in \mathbb{R}[\bm{d}]$.
\item Each component of $\bm{f}(\bm{x},\bm{d})$ is a polynomial in the joint variables $(\bm x, \bm d)$.
\end{enumerate}
\end{assumption}

Under Assumption \ref{assump2}, when searching for a polynomial $v(\bm{a},\bm{x}) \in \mathbb{R}[\bm{x}]$, the uncertain optimization problem \eqref{eq:robust_invariant_set} can be addressed by solving the following semi-definite program \eqref{eq:target_sos}, formulated using a sum-of-squares decomposition of multivariate polynomials.
\begin{align}
\label{eq:target_sos}
    &\textstyle\min_{\bm{a},s_1,s_2,s_3,s_4}
    \quad -\frac{1}{N_r}\sum_{i=1}^{N_r}
    v(\bm{a},\bm{x}'_i)\\
    \text{s.t.}\quad
    &\begin{cases}
    v(\bm{a},\bm{f}(\bm{x},\bm{d}))
    -\gamma v(\bm{a},\bm{x})
    +s_1(\bm{x},\bm{d})g(\bm{x})
    +s_2(\bm{x},\bm{d})p(\bm{d})
    \in\Sigma[\bm{x},\bm{d}],\\
    -v(\bm{a},\bm{x})-\eta_{\rm sos}
    +s_3(\bm{x})\hat g(\bm{x})
    -s_4(\bm{x})g(\bm{x})
    \in\Sigma[\bm{x}],\\
    s_{1}(\bm{x},\bm{d}), s_2(\bm{x},\bm{d}) \in \Sigma[\bm{x},\bm{d}];~s_3(\bm{x}),s_4(\bm{x}) \in \Sigma [\bm{x}];~ \bm{a}\in[-U_a,U_a]^{m_r}.\\
    \end{cases}\notag
\end{align}
Here, $\eta_{\rm sos}>0$ is a small prescribed constant used to enforce the strict negativity condition outside $\mathbb{X}$. Unless otherwise specified, we set $\eta_{\rm sos}=10^{-6}$ throughout this paper. $s_{1}$, $s_2$, $s_3$, and $s_4$ are polynomial multipliers. We additionally impose the coefficient bounds $\bm{a}\in[-U_a,U_a]^{m_r}$ to improve numerical conditioning. After solving \eqref{eq:target_sos}, we obtain the resulting polynomial $v(\bm{a}_v^*,\bm{x})$ and the corresponding robust invariant target set $\mathbb{S}=\{\bm{x} \in \mathbb{X} \mid v(\bm{a}_v^*,\bm{x})\geq0\}$.

We next consider the semi-definite programming implementation of the sample-based optimization problem \eqref{eq:RBF_linear21} for finite-time reach-avoid certification.  A direct semi-definite programming implementation, however, would optimize the polynomial multipliers associated with the unknown coefficients $\bm a$, thereby introducing additional decision variables into the scenario program. To preserve, as closely as possible, the sample-complexity bound in \eqref{MM_so}, which depends only on the dimension $m$ of the unknown parameter $\bm a$, we first determine these multipliers using an auxiliary dataset $\widetilde D^{M_1}=\{\widetilde{\bm d}^{(j)}\}_{j=1}^{M_1}$ with  $\widetilde{\bm d}^{(j)} \overset{\text{i.i.d.}}{\sim} \mathrm{P}_{\bm{d}}$, which is independent of the dataset $D^M=\{\bm d^{(j)}\}_{j=1}^{M}$ in our sampling procedure for PAC characterization.  The auxiliary dataset $\widetilde D^{M_1}=\{\widetilde{\bm d}^{(j)}\}_{j=1}^{M_1}$ is sampled once and then held fixed for all subsequent computations. The corresponding semi-definite program is given below.  
 \begin{align}
\label{eq:rbc_aux_sos}
    &\textstyle\min_{\bm{a},\bm{c}_i,i=1,\ldots,4}
    \quad -\frac{1}{N_o}\sum_{i=1}^{N_o}
    h(\bm{a},\bm{x}'_i)\\
    \text{s.t.}&\begin{cases}
     h(\bm{a}, \bm{f}(\bm{x},\widetilde{\bm d}^{(j)})) - \gamma h(\bm{a}, \bm{x}) + s_1(\bm{c}_1,\bm{x}) g(\bm{x}) + s_2(\bm{c}_2,\bm{x}) v(\bm{a}_v^*,\bm{x}) \in \Sigma [\bm{x}], \\
        -h(\bm{a},\bm{x}) + s_3(\bm{c}_3,\bm{x}) \hat{g}(\bm{x}) - s_4(\bm{c}_4,\bm{x}) g(\bm{x}) \in \Sigma [\bm{x}],\\
        s_1(\bm{c}_1,\bm{x}),s_2(\bm{c}_2,\bm{x}), s_3(\bm{c}_3,\bm{x}),s_4(\bm{c}_4,\bm{x}) \in \Sigma [\bm{x}],\\
         \bm{a}\in[-U_a,U_a]^{m};~\bm c_i\in[-U_s,U_s]^{q_i},~ i = 1,\ldots,4; ~ j=1,\ldots,M_1,
    \end{cases}\notag
\end{align}
where $\bm{c}_i\!\in\!\mathbb{R}^{q_i}$ denotes the vector of unknown coefficients of the polynomial multiplier $s_i$, $i\!=\!1,\!\ldots\!,\!4$, and $U_s>0$ bounds the absolute values of these coefficients.

Let $(\overline{\bm{a}},\overline{\bm{c}}_1,\overline{\bm{c}}_2,\overline{\bm{c}}_3,\overline{\bm{c}}_4)$ be computed via solving \eqref{eq:rbc_aux_sos}. We then fix $\overline{\bm{c}}_1$ and $\overline{\bm{c}}_2$, and solve the following semi-definite program with the dataset $D^M=\{\bm d^{(j)}\}_{j=1}^{M}$:
\begin{align}
\label{eq:rbc2_sos}
    &\textstyle\min_{\bm{a},\kappa,\bm{c}_3,\bm{c}_4}
    \quad -\frac{1}{N_o}\sum_{i=1}^{N_o}
    h(\bm{a},\bm{x}'_i)\\
    \text{s.t.}&\begin{cases}
     h(\bm{a}, \bm{f}(\bm{x},\bm{d}^{(j)})) - \gamma h(\bm{a}, \bm{x}) + \kappa (s_1(\overline{\bm{c}}_1,\bm{x}) g(\bm{x}) + s_2(\overline{\bm{c}}_2,\bm{x}) v(\bm{a}_v^*,\bm{x})) \in \Sigma [\bm{x}], \\
        -h(\bm{a},\bm{x}) + s_3(\bm{c}_3,\bm{x}) \hat{g}(\bm{x}) - s_4(\bm{c}_4,\bm{x}) g(\bm{x}) \in \Sigma[\bm{x}],\\
        s_3(\bm{c}_3,\bm{x}),s_4(\bm{c}_4,\bm{x}) \in \Sigma[\bm{x}],\\
         \bm{a}\in[-U_a,U_a]^{m};~\bm c_3\in[-U_s,U_s]^{q_3};~ \bm c_4\in[-U_s,U_s]^{q_4},\\
         \kappa \in [0, 100];~ j = 1,\ldots,M.
    \end{cases}\notag
\end{align}
It is the scenario program of the following uncertain optimization problem,
\begin{align}
\label{eq:un}
    &\textstyle\min_{\bm{a},\kappa,\bm{c}_3,\bm{c}_4}
    \quad -\frac{1}{N_o}\sum_{i=1}^{N_o}
    h(\bm{a},\bm{x}'_i)\\
    &\begin{cases}
     h(\bm{a},\! \bm{f}(\bm{x},\!\bm{d})) - \gamma h(\bm{a},\! \bm{x}) + \kappa \big(s_1(\overline{\bm{c}}_1,\!\bm{x}) g(\bm{x})+ s_2(\overline{\bm{c}}_2,\!\bm{x}) v(\bm{a}_v^*,\!\bm{x})\big) \in \Sigma [\bm{x}], \forall \bm{d}\in \mathbb{D},\\
        -h(\bm{a},\bm{x}) + s_3(\bm{c}_3,\bm{x}) \hat{g}(\bm{x}) - s_4(\bm{c}_4,\bm{x}) g(\bm{x}) \in \Sigma[\bm{x}],\\
        s_3(\bm{c}_3,\bm{x}),s_4(\bm{c}_4,\bm{x}) \in \Sigma [\bm{x}],\\
         \bm{a}\in[-U_a,U_a]^{m};~\bm c_3\in[-U_s,U_s]^{q_3};~ \bm c_4\in[-U_s,U_s]^{q_4};~\kappa \in [0, 100].
    \end{cases}\notag
\end{align}
Since $g(\bm{x})\leq 0$ on $\mathbb{X}$, $p(\bm{d})\leq 0$ on $\mathbb{D}$, $s_1(\overline{\bm{c}}_1,\bm{x})\geq 0$ and $s_2(\overline{\bm{c}}_2,\bm{x})\geq 0$ for all $\bm{x}\in\mathbb{R}^n$, and $\kappa\geq 0$, the constraint $h(\bm{a},\bm{f}(\bm{x},\bm{d})) -\gamma h(\bm{a},\bm{x})+\kappa\bigl( s_1(\overline{\bm{c}}_1,\bm{x})g(\bm{x}) +s_2(\overline{\bm{c}}_2,\bm{x})v(\bm{a}_v^*,\bm{x}) \bigr) \in\Sigma[\bm{x}]$ implies that $h(\bm{a},\bm{f}(\bm{x},\bm{d})) -\gamma h(\bm{a},\bm{x})\geq 0$ for all $\bm{x}\in\mathbb{X}\setminus\mathbb{S}$ and $\bm{d}\in\mathbb{D}$. Similarly, $ -h(\bm{a},\bm{x}) + s_3(\bm{c}_3,\bm{x}) \hat{g}(\bm{x}) - s_4(\bm{c}_4,\bm{x}) g(\bm{x}) \in \Sigma[\bm{x}]$ implies $h(\bm{a},\bm{x})\leq 0$ for all $\bm{x}\in \overline{\hat{\mathbb{X}}\setminus \mathbb{X}}$. Thus, if $h(\bm{a},\bm{x})$ satisfies the constraints in \eqref{eq:un} and $\{\bm{x}\in \mathbb{X}\mid h(\bm{a},\bm{x})>0\} \neq \emptyset$, $h(\bm{a},\bm{x})$ is an RBGF.

Under Assumption~\ref{as:h}, $\bm a=\bm0$, $\kappa=0$, and $s_3 (\bm x)\equiv s_4 (\bm x)\equiv 0$ form a feasible solution to \eqref{eq:rbc2_sos} for every certification dataset $D^M$. Hence, the optimization \eqref{eq:rbc2_sos} is always feasible. 

The resulting solution satisfies the PAC safety guarantees established in the main text, with $m$ replaced by $m+1$ in the sample-size requirement to account for the additional scalar variable $\kappa$. The multipliers $s_1(\overline{\bm{c}}_1,\bm{x})$ and $s_2(\overline{\bm{c}}_2,\bm{x})$ are fixed using an independent and fixed auxiliary dataset $\widetilde D^{M_1}$ and therefore introduce no additional decision variables during certification. Although $s_3$ and $s_4$ are optimized jointly with $(\bm a,\kappa)$, they and their coefficient bounds appear only in deterministic constraints and thus do not increase the required sample size. These conclusions are formally established in Theorem~\ref{cor:sos_pac}.

\begin{theorem}[PAC Guarantee of the SOS Implementation]
\label{thm:sos_pac}
Let $\epsilon,\delta\in(0,1)$ and $\gamma>1$, and let $\widetilde D^{M_1}$ and $D^M$ be mutually independent datasets of i.i.d. disturbance samples drawn from $\mathbb P_{\bm d}$. Denote the optimal solution to \eqref{eq:rbc2_sos} by $\bm a_{\rm sos}^*(\widetilde D^{M_1},D^M)$. Under Assumption~\ref{as:h}, if
\begin{equation}
\label{MM_so_sos}
M\geq\frac{2}{\epsilon}
\left(\ln\frac{1}{\delta}+m+1\right),
\end{equation}
then the following hold:
\begin{enumerate}
    \item[(i)] For every $\bm{x}\in \mathbb{S}=\{\bm{x}\! \in\! \mathbb{X} \mid v(\bm{a}_v^*,\bm{x})\geq0\}$, $\mathrm{P}_{\pi}\big[\bm{\phi}_{\pi}^{\bm{x}}(j)\in\mathbb{X}\ \text{for all }j\in\mathbb{N}\big]=1$.
    \item[(ii)] With probability at least $1-\delta$ over $D^{M}$, if the certification procedure accepts, i.e., if $\mathbb{X}_{s,\beta}(\widetilde D^{M_1},D^{M})=\big\{\bm{x}\in\mathbb{X}\mid h(\bm{a}_{\rm sos}^*(\widetilde D^{M_1},D^{M}),\bm{x})\ge\beta\big\}\neq\emptyset$, then
    \begin{equation}
    \label{eq:pac_inf}
    \forall\bm{x}\in\mathbb{X}_{s,\beta}(\widetilde D^{M_1},D^{M})\setminus\mathbb{S}:\quad
    \mathrm{P}_{\pi}\big[\bm{\phi}_{\pi}^{\bm{x}}(j)\in\mathbb{X}\ \text{for all }j\in\mathbb{N}\big]\ge(1-\epsilon)^K,
    \end{equation}
    where $H=\sup_{\bm{x}\in\mathbb{X}\setminus\mathbb{S}}h(\bm{a}_{\rm sos}^*(\widetilde D^{M_1},D^{M}),\bm{x})$ and $
    K=\Big\lfloor\log_{\gamma}\frac{H}{\beta}\Big\rfloor+1$.
\end{enumerate}
\end{theorem}
\begin{proof}
The proof is shown in Appendix \ref{sec:pro}. \qed
\end{proof}
\section{Illustrative Examples}
\label{sec:ex}

We demonstrate the proposed semi-definite programming approaches on several numerical examples.

\noindent\textbf{Baselines.} We evaluate the proposed framework against three baselines, each isolating one ingredient of the method. The first solves
the semi-definite program~\eqref{eq:target_sos} and therefore certifies only the robust invariant set $\mathbb{S}$: every $\bm{x}\in\mathbb{S}$ is safe with probability one, but no initial state outside $\mathbb{S}$ is certified. Comparing against it isolates the benefit of the reach-avoid step, which extends certification from $\mathbb{S}$ to the larger set $\mathbb{X}_{s,\beta}\cup\mathbb{S}$. The second baseline solves the semi-definite program \eqref{eq:rbc1_sos}, which encodes the worst-case RBGF program \eqref{eq:RBF_linear1} via a sum-of-squares decomposition: the barrier condition is required to hold for every disturbance in $\mathbb{D}$, in contrast to the sampled counterpart \eqref{eq:RBF_linear21}; comparing against it isolates the benefit of replacing a worst-case requirement over $\mathbb{D}$ by one over finitely many samples. The third baseline is the robust-barrier-function-based PAC finite-time safety certification of \cite{wu2026pac}, which addresses the same class of stochastic systems with unknown disturbance distributions but provides guarantees only over a finite horizon; it isolates the benefit of the infinite-horizon decomposition. Together, the three baselines separate the contributions of the reach-avoid step, of the sampled relaxation, and of the infinite-horizon reduction.

In all experiments, we set $\delta=0.001$, so that all guarantees hold with confidence at least $99.9\%$ by Theorem~\ref{thm:sos_pac}, and for each $\epsilon$ we take the smallest $M$ satisfying the PAC sample-size requirement. For a fair comparison, the polynomial degrees of the barrier templates are kept the same across the robust invariant-set computation, the worst-case RBGF formulation, the proposed method, and the baseline of \cite{wu2026pac}. Set sizes are compared by Monte Carlo: we draw $10^{7}$ states uniformly from the safe set $\mathbb{X}$ once and, for every $\mathbb{Y}\subseteq\mathbb{X}$, report $\widehat{V}_{\mathbb{Y}}=N_{\mathbb{Y}}/10^{7}$, where $N_{\mathbb{Y}}$ is the number of sampled states in $\mathbb{Y}$. 
Examples~\ref{ex:1} and~\ref{ex:lotka} are examined in detail below; additional results are summarized in Table~\ref{tab:exp}.

\begin{example}
\label{ex:1}
    Consider the system given by
\begin{equation*}
    \begin{cases}
x(k+1)=0.5x(k)+(0.4+x^2(k)+y^2(k))x(k)d(k),\\
y(k+1)=0.5y(k)+(0.4+x^2(k)+y^2(k))y(k)d(k),
\end{cases}
\end{equation*}
where $\mathbb{X}=\{\,(x,y)^{\top}\mid x^2+y^2 \leq 1\,\}$. Besides, $d(k)$ follows a symmetric Beta distribution scaled to $[-1, 1]$, i.e., $d(k)=2 Z(k)-1$ with $Z(k) \sim\text{Beta}(100,100)$.

\begin{figure}[!htbp]
    \centering
    \subfigure[]{\includegraphics[width=0.32\linewidth]{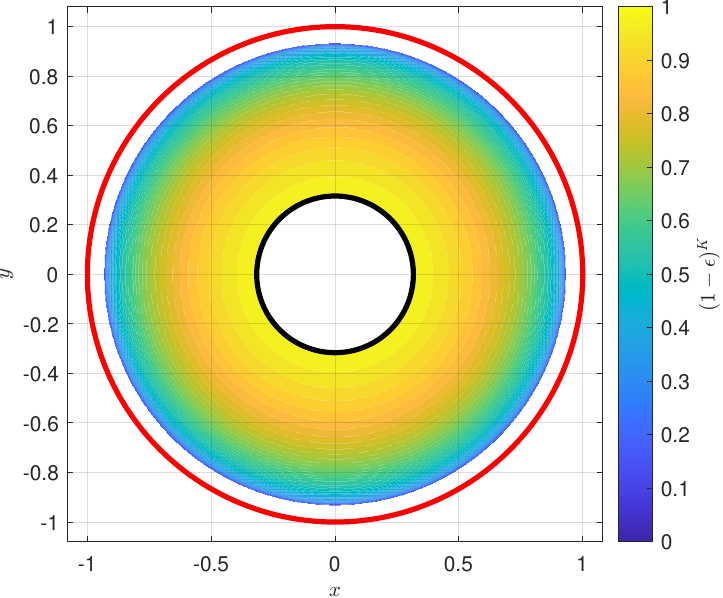}\label{fig:exp1_beta}}\hfill
    \subfigure[]{\includegraphics[width=0.27\linewidth]{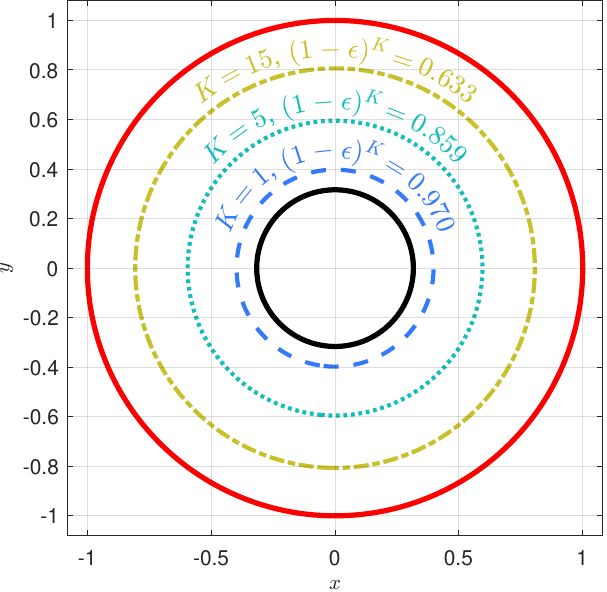}\label{fig:exp1_K}}\hfill
    \subfigure[]{\includegraphics[width=0.37\linewidth]{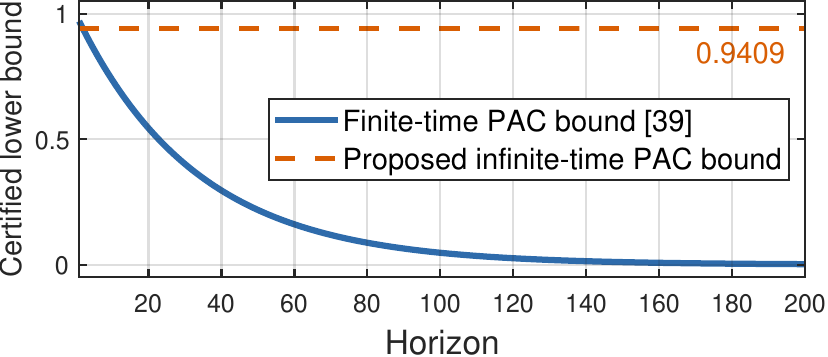}\label{fig:prob_k}}
    \caption{PAC safety certification for Example 1. (a) Certified sets obtained by the proposed method for different probability lower bounds. (b) Three representative certified sets. In (a) and (b), the \textcolor{red}{red} and black curves mark the boundaries of the safe set $\mathbb{X}$ and the robust invariant set $\mathbb{S}$, respectively. (c) Comparison of finite- and infinite-horizon PAC safety bounds for the initial state $(0.45,0)$.}
    \label{fig:exp1}
\end{figure}

In this example, the maximal robust invariant set can be obtained analytically and has relative volume $0.10$, as enclosed by the black curve in Fig.~\ref{fig:exp1_beta}. Since this set is already the maximal robust invariant set, using it as the target in the semi-definite program~\eqref{eq:rbc1_sos} yields an empty set $\mathbb{X}_s$, indicating that the worst-case RBGF method cannot certify any additional states.
We then solve the semi-definite programs \eqref{eq:rbc_aux_sos} and \eqref{eq:rbc2_sos} with a degree-$4$ polynomial $h(\bm{a},\bm{x})$ and $\epsilon=0.03$. The computation takes $179$~s, and the certified set $\mathbb{X}_{s,\beta}$ has relative volume $0.869$, compared with $0.10$ for the robust invariant set alone (Fig.~\ref{fig:exp1}). Thus, even when the target set is already maximal under worst-case robustness, our method delivers nontrivial infinite-time PAC guarantees over a substantially larger portion of the safe set, using only finitely many disturbance samples.

The certification results are visualized in Fig. \ref{fig:exp1}. Fig. \ref{fig:exp1}(a) shows the nested family of certified sets obtained by varying $\beta$, with contours reporting the uniform lower bound $(1-\epsilon)^{K}$ guaranteed by Theorem~\ref{thm:sos_pac} on each set; Fig.~\ref{fig:exp1}(b) highlights three representative instances with horizons $K=1$, $5$, and $15$, whose bounds are $0.97$, $0.86$, and $0.63$. Note that Theorem \ref{thm:sos_pac} states a single uniform bound on each $\mathbb{X}_{s,\beta}$, equal to its worst-case value over the set and attained at states with $h_{\rm sos}(\bm{a}^*,\bm{x})=\beta$. The variation visible across the figure therefore comes from comparing nested sets, not from a per-state bound: by Remark \ref{rem:sharpen}, a state lying in a smaller set $\mathbb{X}_{s,\beta'}$ with $\beta'>\beta$ is certified by the stronger guarantee $(1-\epsilon)^{K'}$, with $K'\le K$, and in particular the bound available at $\bm{x}$ is obtained by taking $\beta'=h_{\rm sos}(\bm{a}^*,\bm{x})$. The framework thus provides an explicit tradeoff between the size of the certified region and the strength of the infinite-time safety guarantee—tunable through $\beta$ from a single synthesized RBGF, without re-solving the optimization—and, in this nested sense, the certification reflects differences in the safety levels of different states, while each individual bound remains a rigorous uniform guarantee on its own certified set.

Under the same setting ($\epsilon=0.03$), the finite-time PAC method of \cite{wu2026pac} certifies a region covering almost the entire safe set in $136.9$~s. Its guarantee, however, is confined to a prescribed horizon: the $K$-step safety probability is bounded below by $(1-\epsilon)^{K}$, which decays exponentially in $K$. In the proposed framework, $K$ counts only the steps needed to reach the robust invariant set, and the resulting bound applies over the entire infinite horizon. Fig.~\ref{fig:prob_k} illustrates this for the initial state $(0.45,0)$: the proposed framework certifies an infinite-horizon lower bound of $0.9409$, whereas the bound of \cite{wu2026pac} decreases with the prescribed horizon and approaches zero beyond $100$ steps. A second difference concerns granularity. The method of \cite{wu2026pac} assigns the same bound to every state in its certified set, whereas the proposed framework, through the nested family of certified sets, yields stronger guarantees for states from which the robust invariant set is reached within fewer steps.
\end{example}

\begin{example}
\label{ex:lotka}
    Consider the following discrete-time Lotka–Volterra model \cite{lotka2002contribution} for two interacting species
\begin{equation*}
    \begin{cases}
x(k+1)=rx(k)-ay(k)x(k),\\
y(k+1)=(-0.5+d(k))y(k)+acy(k)x(k),
\end{cases}
\end{equation*}
where $r=0.5$, $a=c=1$, and $\mathbb{X}=\{\,(x,y)^{\top}\mid x^2+y^2 \leq 1\,\}$. Besides, $d(k)$ follows a truncated Student's $t$-distribution defined on $[-0.5,0.5]$, with zero location, scale parameter $0.03$, and $5$ degrees of freedom.

In this example, all polynomial templates are set to degree $6$ and $\epsilon=0.03$. The robust invariant set, enclosed by the black curve in Fig.~\ref{fig:lotka}, has relative volume $0.656$. Solving the worst-case RBGF program \eqref{eq:rbc1_sos} with this set as the target yields a certified region of volume $0.798$ (Fig.~\ref{fig:lotka_robust}) that is itself robustly invariant—a moderate enlargement. This implies that the initially computed  robust invariant set is not the maximal one. The proposed sample-based program \eqref{eq:rbc2_sos} instead certifies a substantially larger region of volume $0.937$ (Fig.~\ref{fig:lotka_beta_1}), enlarging the certified region relative to worst-case formulations at the cost of replacing a probability-one guarantee, valid for every disturbance in $\mathbb{D}$, with a PAC guarantee. Fig.~\ref{fig:lotka_finite} also shows three trajectories from the initial state $(-0.5,-0.6)$ for different disturbance realizations, some remaining in the safe set and others leaving it. This coexistence of safe and unsafe trajectories illustrates why probabilistic certification can cover states that worst-case analysis excludes. 
For comparison, the finite-time PAC method of \cite{wu2026pac} certifies a region of volume $0.911$ (Fig.~\ref{fig:lotka_finite}), guaranteeing a $K$-step safety probability of at least $0.97^{K}$, whereas our method certifies a slightly larger set with infinite-time safety. 

As an empirical sanity check, we perform independent Monte Carlo simulations, with the sampling details given in Appendix~\ref{sec:app_exp}. Across all sampled states, the observed one-step satisfaction probabilities exceed $0.996$, consistently above the theoretical lower bound $1-\epsilon=0.97$, in agreement with Lemma~\ref{lemma:robust_s}. Moreover, empirical safety rates from long-horizon simulations over $10^3$ steps exceed the corresponding certified infinite-time lower bounds. These observations are consistent with the theoretical guarantees.

\begin{figure*}[!h]
    \vspace{-10pt}
    \centering
     \subfigure[Ours]{
    \includegraphics[width=2.95cm,height=2.55cm]{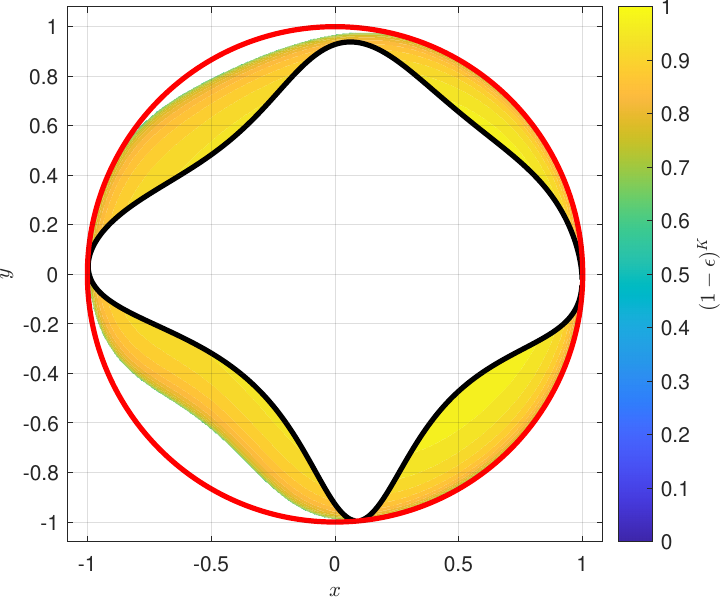}
    \label{fig:lotka_beta_1}
    }
    \subfigure[Worst-case RBGF formulation]{
    \includegraphics[width=2.55cm,height=2.55cm]{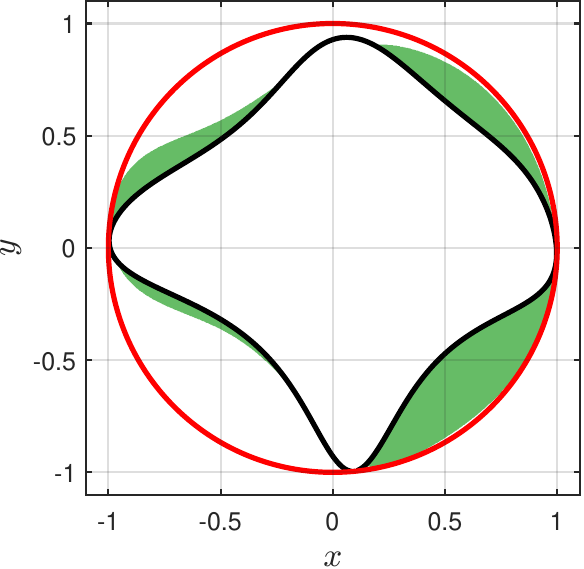}
    \label{fig:lotka_robust}
    }
    \subfigure[Ours (small target set)]{
    \includegraphics[width=2.95cm,height=2.55cm]{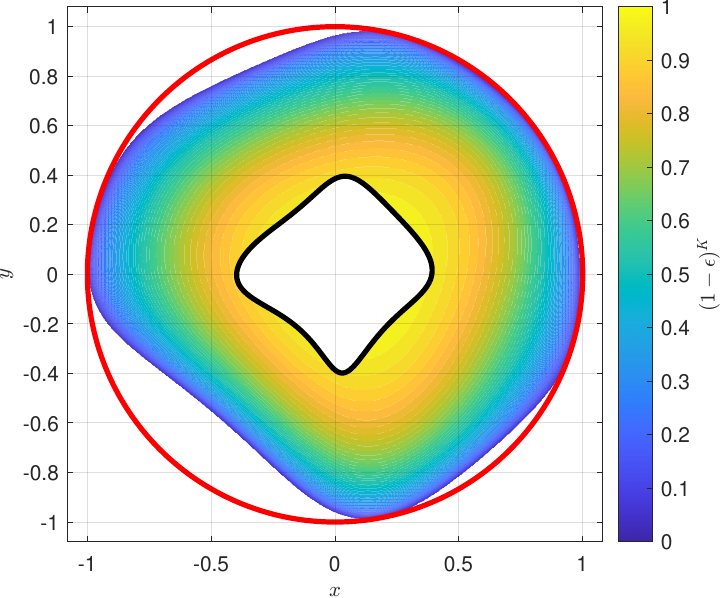}
    \label{fig:lotka_beta_2}
    }
    \subfigure[Finite-time \cite{wu2026pac} and Trajectories]{
    \includegraphics[width=2.55cm,height=2.55cm]{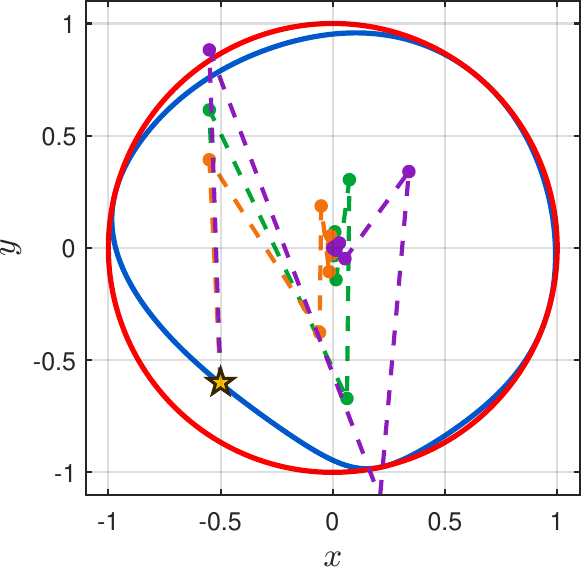}
    \label{fig:lotka_finite}
    }
    \caption{Results for Example~\ref{ex:lotka}. \textcolor{red}{Red} and black curves represent the boundaries of the safe set $\mathbb{X}$ and the robust invariant set $\mathbb{S}$.}
    \label{fig:lotka}
    \vspace{-10pt}
\end{figure*}

\noindent\textbf{Effect of the robust invariant set.} By Theorem~\ref{thm:ris}, if no nonempty robust invariant set exists within $\mathbb{X}$, then the infinite-time safety probability is zero for every initial state in $\mathbb{X}$, precluding any nontrivial infinite-time safety guarantee. In practice, however, one may only have access to a robust invariant set $\mathbb{S}$, without knowing whether it is maximal or how conservative it is. We therefore investigate how the choice of $\mathbb{S}$ affects the proposed certification. Using a substantially smaller robust invariant set with relative volume $0.116$ yields the results in Fig.~\ref{fig:lotka_beta_2}, where the largest certified region has a relative volume of $0.891$. A smaller target may require more steps for states outside $\mathbb{S}$ to reach it, potentially weakening the resulting probability lower bounds. Nevertheless, the proposed method does not require a maximal robust invariant set and can still certify a region substantially larger than $\mathbb{S}$ itself. This example demonstrates that the method can mitigate the conservatism associated with a suboptimal choice of invariant target while providing PAC safety guarantees.

\end{example}

\paragraph{Results across benchmarks.} To further evaluate the proposed method, we conduct experiments on nine benchmark systems, with the results summarized in Table~\ref{tab:exp}. The reported values are averaged over three random seeds, while the corresponding standard deviations are provided in Appendix Tables~\ref{tab:exp_volume} and~\ref{tab:exp_time}. For the proposed method, $\widehat V_{\max}$ denotes the relative volume of the largest certified region, while $\widehat V_{0.8}$ and $\widehat V_{0.5}$ denote the relative volumes of the certified regions for which the infinite-horizon safety probability is at least $0.8$ and $0.5$, respectively. For Examples~\ref{ex:1} and~\ref{ex:6d}, the maximal robust invariant set can be obtained analytically from the system structure, so no numerical invariant-set computation is required and the corresponding computation time is reported as $0$.

Across all examples, the proposed method consistently enlarges the certified region compared with both the robust invariant set and the worst-case RBGF baseline. In particular, even when an infinite-horizon safety probability is required to be no less than $0.8$, the proposed method yields a larger certified region than both worst-case baselines in all examples. The largest certified regions are also comparable to those obtained by the finite-time PAC baseline~\cite{wu2026pac} in most cases, while providing infinite-time rather than finite-horizon safety guarantees. Similar improvements are observed across all considered systems, including the six-dimensional system in Example~\ref{ex:6d}. Overall, these results demonstrate that exploiting sampled disturbance information can substantially reduce the conservatism of worst-case certification while providing rigorous PAC guarantees for infinite-time safety.
In Appendix~\ref{sec:app_ex_results}, we examine the effects of $\epsilon$, the polynomial degree, and $\gamma$ and analyze our method under different disturbance distributions.

\begin{table*}[!htb]
\caption{Overall experimental results (Time is reported in seconds).}
\setlength{\tabcolsep}{0.9mm}{
\begin{tabular}{ccccccccccc}
\hline
\multirow{2.5}{*}{Ex}
& \multicolumn{2}{c}{Robust Invariant Set}
& \multicolumn{2}{c}{Worst-case RBGF}
& \multicolumn{4}{c}{Ours}
& \multicolumn{2}{c}{Finite Time~\cite{wu2026pac}} \\
\cmidrule(lr){2-3}
\cmidrule(lr){4-5}
\cmidrule(lr){6-9}
\cmidrule(lr){10-11}
& Time & $\widehat V$
& Time & $\widehat V$
& Time & $\widehat V_{\max}$ & $\widehat V_{0.8}$ & $\widehat V_{0.5}$
& Time & $\widehat V$ \\
\hline
1 & 0.0 & $0.100$ & 1.0 & $0.100$ & 247 & $0.856$ & $0.428$ & $0.742$ & 220 & $1.000$ \\
2 & 1.2 & $0.669$ & 2.3 & $0.786$ & 621 & $0.922$ & $0.889$ & $0.922$ & 251 & $0.898$ \\
3 & 2.4 & $0.519$ & 6.9 & $0.519$ & 918 & $1.000$ & $0.904$ & $0.998$ & 468 & $0.999$ \\
4 & 1.7 & $0.299$ & 3.1 & $0.299$ & 975 & $0.637$ & $0.570$ & $0.636$ & 517 & $0.694$ \\
5 & 0.9 & $0.650$ & 1.2 & $0.650$ & 98  & $0.903$ & $0.890$ & $0.903$ & 69  & $0.927$ \\
6 & 1.6 & $0.517$ & 3.0 & $0.888$ & 489 & $1.000$ & $0.982$ & $1.000$ & 505 & $1.000$ \\
\ref{ex:rotor3d} & 0.5 & $0.653$ & 0.7 & $0.790$ & 23 & $0.976$ & $0.964$ & $0.976$ & 25 & $1.000$ \\
\ref{ex:shg4d} & 0.6 & $0.644$ & 0.7 & $0.648$ & 73 & $0.988$ & $0.972$ & $0.988$ & 62 & $1.000$ \\
\ref{ex:6d} & 0.0 & $0.065$ & 1.2 & $0.065$ & 519 & $1.000$ & $0.191$ & $0.725$ & 589 & $1.000$ \\
\hline
\end{tabular}}
\label{tab:exp}
\vspace{-7pt}
\end{table*}

\section{Conclusion}
\label{sec:con}

This work studied PAC infinite-time safety guarantees for discrete-time stochastic systems with unknown disturbance distributions. By establishing the equivalence between positive infinite-time safety probability for some initial state and the existence of a robust invariant set within the safe set, we reduce infinite-time safety certification to two subproblems: robust invariant-set computation and PAC finite-time reach-avoid certification. When the system is polynomial, both subproblems are encoded as semi-definite programs via sum-of-squares decompositions, and combining them with the scenario approach yields infinite-time safety guarantees from finitely many disturbance samples. Numerical experiments confirmed the effectiveness of the framework.

Future work will study PAC infinite-time safety certification using stochastic barrier functions. Also, we aim to extend the proposed approaches to PAC-style verification for probabilistic programs and quantum computations.

 \bibliographystyle{unsrt}
\bibliography{ref}

\section*{Appendix}
\setcounter{secnumdepth}{2}
\setcounter{section}{0}
\renewcommand\thesubsection{\thesection.\arabic{subsection}}
\renewcommand\thesection{\Alph{section}}
\label{sec:appendix}

 \section{Proofs in the Paper}
 \label{sec:pro}
 \setcounter{lemma}{0}
\setcounter{theorem}{0}
 \begin{lemma}[Equivalence Between Almost-Sure Invariance and Positive Safety Probability]
The following statements are equivalent.
\begin{enumerate}
    \item[(1)] There exists a nonempty set $\mathbb{S}\subseteq \mathbb{X}$ such that
    \[\mathrm{P}_{\pi}\Big[ \bigwedge_{j=0}^{\infty} \bm{\phi}^{\bm{x}}_{\pi}(j) \in \mathbb{S}\Big]=1, \qquad \forall \bm{x}\in \mathbb{S}.\]
    \item[(2)] There exists a state $\bm{x}\in \mathbb{X}$ satisfying
    \[\mathrm{P}_{\pi}\Big[\bigwedge_{j=1}^{\infty} \bm{\phi}^{\bm{x}}_{\pi}(j)\in \mathbb{X}\Big]>0.\]
\end{enumerate}
\end{lemma}

\begin{proof}
$(1)\Rightarrow(2)$: take $\bm{x}\in\mathbb{S}$. Since $\mathbb{S}\subseteq\mathbb{X}$ and $\bm{\phi}^{\bm{x}}_{\pi}(0)=\bm{x}$,
\[
\mathrm{P}_{\pi}\Big[\bigwedge_{j=1}^{\infty}\bm{\phi}^{\bm{x}}_{\pi}(j)\in\mathbb{X}\Big]
\ge \mathrm{P}_{\pi}\Big[\bigwedge_{j=1}^{\infty}\bm{\phi}^{\bm{x}}_{\pi}(j)\in\mathbb{S}\Big]=1>0 .
\]

$(2)\Rightarrow(1)$: define the infinite-horizon safety probability 
\[
V(\bm{x}):=\mathrm{P}_{\pi}\Big[\bigwedge_{k=0}^{\infty}\bm{\phi}^{\bm{x}}_{\pi}(k)\in\mathbb{X}\Big]
=\mathrm{E}_{\pi}\Big[\prod_{k=0}^{\infty}1_{\mathbb{X}}\big(\bm{\phi}^{\bm{x}}_{\pi}(k)\big)\Big],
\qquad \bm{x}\in\mathbb{R}^n .
\]
For $\bm{x}\in\mathbb{X}$ this coincides with the probability in (2), because $\bm{\phi}^{\bm{x}}_{\pi}(0)=\bm{x}\in\mathbb{X}$; hence $V(\bm{x}_0)>0$ for some $\bm{x}_0\in\mathbb{X}$ by hypothesis. Since $V(\bm{x})=0$ for $\bm{x}\in\mathbb{R}^n\setminus\mathbb{X}$, the Markov property gives, for every $\bm{x}\in\mathbb{X}$,
\[
V(\bm{x})=\mathrm{E}_{\bm{d}}\big[1_{\mathbb{X}}\big(\bm{f}(\bm{x},\bm{d})\big)\,V\big(\bm{f}(\bm{x},\bm{d})\big)\big].
\]
By Lemma \ref{upper_con} (shown below), $V$ is upper semicontinuous on the compact set $\mathbb{X}$ and therefore attains its maximum
\[
m:=\max_{\bm{x}\in\mathbb{X}}V(\bm{x})>0 .
\]
Let $\mathbb{S}:=\{\bm{x}\in\mathbb{X}\mid V(\bm{x})=m\}$, which is nonempty and closed in $\mathbb{X}$. Fix $\bm{x}\in\mathbb{S}$ and define the random variable
\[
Y:=1_{\mathbb{X}}\big(\bm{f}(\bm{x},\bm{d})\big)\,V\big(\bm{f}(\bm{x},\bm{d})\big).
\]
Since $V\le m$ on $\mathbb{X}$, we have $Y\le m$ almost surely: if $\bm{f}(\bm{x},\bm{d})\in\mathbb{X}$ then $Y=V(\bm{f}(\bm{x},\bm{d}))\le m$, and otherwise $Y=0\le m$. On the other hand, applying the recursion at $\bm{x}\in\mathbb{S}\subseteq\mathbb{X}$ gives
\[
\mathrm{E}_{\bm{d}}[Y]=\mathrm{E}_{\bm{d}}\big[1_{\mathbb{X}}\big(\bm{f}(\bm{x},\bm{d})\big)V\big(\bm{f}(\bm{x},\bm{d})\big)\big]=V(\bm{x})=m .
\]
Hence $m-Y\ge0$ almost surely and $\mathrm{E}_{\bm{d}}[m-Y]=0$, so $m-Y=0$ $\mathrm{P}_{\bm{d}}$-almost surely. We now decompose
\[
m-Y
=
\underbrace{m\Big(1-1_{\mathbb{X}}\big(\bm{f}(\bm{x},\bm{d})\big)\Big)}_{=:A}
+
\underbrace{1_{\mathbb{X}}\big(\bm{f}(\bm{x},\bm{d})\big)\Big(m-V\big(\bm{f}(\bm{x},\bm{d})\big)\Big)}_{=:B},
\]
as is readily verified by expansion. Both $A$ and $B$ are nonnegative: $A\ge0$ because $m>0$ and $1_{\mathbb{X}}\le1$, and $B\ge0$ because $V\le m$ on $\mathbb{X}$. Since $A+B=m-Y=0$ almost surely, each summand vanishes almost surely.

From $A=0$ almost surely and $m>0$ we obtain
\[
1_{\mathbb{X}}\big(\bm{f}(\bm{x},\bm{d})\big)=1 \quad \mathrm{P}_{\bm{d}}\text{-almost surely},
\qquad\text{i.e.}\qquad
\mathrm{P}_{\bm{d}}\big[\bm{f}(\bm{x},\bm{d})\in\mathbb{X}\big]=1 .
\]
Combining this with $B=0$ almost surely yields $m-V(\bm{f}(\bm{x},\bm{d}))=0$ $\mathrm{P}_{\bm{d}}$-almost surely, that is, $V(\bm{f}(\bm{x},\bm{d}))=m$ for $\mathrm{P}_{\bm{d}}$-almost every $\bm{d}$. Equivalently,
\[
\mathrm{P}_{\bm{d}}\big[\bm{f}(\bm{x},\bm{d})\in\mathbb{S}\big]=1 .
\]

We next prove by induction that $\mathrm{P}_{\pi}[\bm{\phi}^{\bm{x}}_{\pi}(k)\in\mathbb{S}]=1$ for all $k\ge0$. The claim holds for $k=0$ because $\bm{\phi}^{\bm{x}}_{\pi}(0)=\bm{x}\in\mathbb{S}$. Assume it holds for some $k\ge0$. Define the Borel function
\[
g(\bm{y}):=\mathrm{P}_{\bm{d}}\big[\bm{f}(\bm{y},\bm{d})\in\mathbb{S}\big].
\]
Since $\{\bm{\phi}^{\bm{x}}_{\pi}(k)\}$ is a time-homogeneous Markov chain with transition kernel $P(\bm{y},A)=\mathrm{P}_{\bm{d}}[\bm{f}(\bm{y},\bm{d})\in A]$, the Markov property gives
\[
\mathrm{P}_{\pi}\big[\bm{f}(\bm{\phi}^{\bm{x}}_{\pi}(k),\bm{d}(k))\in\mathbb{S}\,\big|\,\sigma(\bm{\phi}^{\bm{x}}_{\pi}(k))\big]
=g\big(\bm{\phi}^{\bm{x}}_{\pi}(k)\big)
\quad \mathrm{P}_{\pi}\text{-almost surely}.
\]
Applying the tower property to the indicator of $\{\bm{\phi}^{\bm{x}}_{\pi}(k+1)\in\mathbb{S}\}$ and substituting the identity above,
\[
\mathrm{P}_{\pi}\big[\bm{\phi}^{\bm{x}}_{\pi}(k+1)\in\mathbb{S}\big]
=\mathrm{P}_{\pi}\big[\bm{f}(\bm{\phi}^{\bm{x}}_{\pi}(k),\bm{d}(k))\in\mathbb{S}\big]
=\mathrm{E}_{\pi}\big[g\big(\bm{\phi}^{\bm{x}}_{\pi}(k)\big)\big]=1,
\]
where the last equality uses $\bm{\phi}^{\bm{x}}_{\pi}(k)\in\mathbb{S}$ almost surely and $g(\bm{y})=1$ for every $\bm{y}\in\mathbb{S}$.
Since $\mathrm{P}_{\bm{d}}[\bm{f}(\bm{y},\bm{d})\in\mathbb{S}]=1$ for every $\bm{y}\in\mathbb{S}$ and $\bm{\phi}^{\bm{x}}_{\pi}(k)\in\mathbb{S}$ almost surely, the integrand equals one almost surely, so $\mathrm{P}_{\pi}[\bm{\phi}^{\bm{x}}_{\pi}(k+1)\in\mathbb{S}]=1$. This completes the induction.

Taking the countable intersection of these probability-one events,
\[
\mathrm{P}_{\pi}\Big[\bigwedge_{k=1}^{\infty}\bm{\phi}^{\bm{x}}_{\pi}(k)\in\mathbb{S}\Big]=1,\qquad \forall \bm{x}\in\mathbb{S}.
\]
Together with $\bm{\phi}^{\bm{x}}_{\pi}(0)=\bm{x}\in\mathbb{S}$, this yields
\[
\mathrm{P}_{\pi}\Big[\bigwedge_{k=0}^{\infty}\bm{\phi}^{\bm{x}}_{\pi}(k)\in\mathbb{S}\Big]=1,\qquad \forall \bm{x}\in\mathbb{S},
\]
showing that $\mathbb{S}$ is an almost-sure probabilistic invariant set, i.e.\ (1) holds. \qed
\end{proof}

The proof of Lemma ~\ref{thm:aps} relies on the infinite-horizon safety probability attaining its maximum over the compact safe set, which in turn follows from its upper semicontinuity. The following lemma establishes this regularity property.

 \setcounter{lemma}{3}
\begin{lemma}
 \label{upper_con}
  The  infinite-horizon safety probability
\[
V(\bm{x}):=\mathrm{P}_{\pi}\Big[\bigwedge_{k=0}^{\infty}\bm{\phi}_{\pi}^{\bm{x}}(k)\in\mathbb{X}\Big]
=\mathrm{E}_{\pi}\Big[\prod_{k=0}^{\infty}1_{\mathbb{X}}\big(\bm{\phi}_{\pi}^{\bm{x}}(k)\big)\Big]
\]
is upper semicontinuous on $\mathbb{R}^n$, and hence on $\mathbb{X}$.
\end{lemma}

\begin{proof}
We approximate $V$ by finite-horizon safety probabilities and show that the approximation is upper semicontinuous at every finite horizon. For $N\ge0$, define
\[
V_N(\bm{x}):=\mathrm{P}_{\pi}\Big[\bigwedge_{k=0}^{N}\bm{\phi}_{\pi}^{\bm{x}}(k)\in\mathbb{X}\Big],\qquad \bm{x}\in\mathbb{R}^n,
\]
with $V_N(\bm{x})=0$ for $\bm{x}\notin\mathbb{X}$ by construction. The events are nested in $N$,
\[
\Big\{\bm{\phi}_{\pi}^{\bm{x}}(k)\in\mathbb{X},\ k=0,\dots,N+1\Big\}\subseteq\Big\{\bm{\phi}_{\pi}^{\bm{x}}(k)\in\mathbb{X},\ k=0,\dots,N\Big\},
\]
so $V_N(\bm{x})\downarrow V(\bm{x})$ pointwise as $N\to\infty$; by continuity of the probability measure from above,
\begin{equation}\label{eq:Vinf}
V(\bm{x})=\lim_{N\to\infty}V_N(\bm{x})=\inf_{N\ge0}V_N(\bm{x}),\qquad \bm{x}\in\mathbb{R}^n.
\end{equation}

We now show, by induction on $N$, that each $V_N$ is upper semicontinuous on $\mathbb{R}^n$.

\medskip
\noindent\textbf{Base case ($N=0$).} $V_0=1_{\mathbb{X}}$, which is upper semicontinuous on $\mathbb{R}^n$ because $\mathbb{X}$ is compact, hence closed: for $\bm{x}\in\mathbb{X}$ we have $\limsup_{\bm{y}\to\bm{x}}1_{\mathbb{X}}(\bm{y})\le 1=1_{\mathbb{X}}(\bm{x})$, and for $\bm{x}\notin\mathbb{X}$ closedness yields a neighborhood on which $1_{\mathbb{X}}\equiv 0$.

\medskip
\noindent\textbf{Induction step ($N\ge1$).} Assume that $V_{N-1}$ is upper semicontinuous on $\mathbb{R}^n$. By the Markov property, for every $\bm{x}\in\mathbb{R}^n$,
\begin{equation}\label{eq:recursion}
V_N(\bm{x})=1_{\mathbb{X}}(\bm{x})\,\mathrm{E}_{\bm{d}}\Big[1_{\mathbb{X}}\big(\bm{f}(\bm{x},\bm{d})\big)\,V_{N-1}\big(\bm{f}(\bm{x},\bm{d})\big)\Big].
\end{equation}
Indeed, for $\bm{x}\in\mathbb{X}$ this is the Markov property together with the definition of $V_{N-1}$, and for $\bm{x}\notin\mathbb{X}$ both sides vanish. Define
\[
G(\bm{x},\bm{d}):=1_{\mathbb{X}}\big(\bm{f}(\bm{x},\bm{d})\big)\,V_{N-1}\big(\bm{f}(\bm{x},\bm{d})\big),\qquad (\bm{x},\bm{d})\in\mathbb{R}^n\times\mathbb{D}.
\]

 Since $\bm{f}$ is continuous and $\mathbb{X}$ is closed, $1_{\mathbb{X}}\circ\bm{f}$ is upper semicontinuous as a composition of a continuous map with an upper semicontinuous function; by the induction hypothesis, $V_{N-1}\circ\bm{f}$ is upper semicontinuous. Both factors are nonnegative and bounded by $1$, so their product is upper semicontinuous and takes values in $[0,1]$. This proves the claim.

Now let $\{\bm{x}_m\}_{m\ge1}\subseteq\mathbb{R}^n$ with $\bm{x}_m\to\bm{x}$. By the claim, for every fixed $\bm{d}\in\mathbb{D}$,
\[
\limsup_{m\to\infty}G(\bm{x}_m,\bm{d})\le G(\bm{x},\bm{d}).
\]
Since $0\le G\le1$, the reverse Fatou lemma applies to the bounded sequence $\{G(\bm{x}_m,\cdot)\}_{m\ge1}$:
\begin{equation*}\label{eq:revfatou}
\limsup_{m\to\infty}\int_{\mathbb{D}}G(\bm{x}_m,\bm{d})\,\mathrm{P}_{\bm{d}}(d\bm{d})
\le\int_{\mathbb{D}}\limsup_{m\to\infty}G(\bm{x}_m,\bm{d})\,\mathrm{P}_{\bm{d}}(d\bm{d})
\le\int_{\mathbb{D}}G(\bm{x},\bm{d})\,\mathrm{P}_{\bm{d}}(d\bm{d}).
\end{equation*}
(Equivalently, apply Fatou's lemma to the nonnegative sequence $1-G(\bm{x}_m,\bm{d})$ and use $\liminf_m(1-G(\bm{x}_m,\bm{d}))=1-\limsup_m G(\bm{x}_m,\bm{d})$.) Consequently, the map
\[
\bm{x}\mapsto H(\bm{x}):=\mathrm{E}_{\bm{d}}\big[G(\bm{x},\bm{d})\big]=\int_{\mathbb{D}}G(\bm{x},\bm{d})\,\mathrm{P}_{\bm{d}}(d\bm{d})
\]
is upper semicontinuous on $\mathbb{R}^n$. Since $1_{\mathbb{X}}$ is upper semicontinuous and both $1_{\mathbb{X}}$ and $H$ are nonnegative, the product $V_N=1_{\mathbb{X}}\cdot H$ is upper semicontinuous on $\mathbb{R}^n$ by \eqref{eq:recursion}. This completes the induction.

\medskip
Finally, by \eqref{eq:Vinf}, $V$ is the pointwise infimum of the family $\{V_N\}_{N\ge0}$ of upper semicontinuous functions. For every $\alpha\in\mathbb{R}$,
\[
\Big\{\bm{x}\in\mathbb{R}^n \mid  V(\bm{x})<\alpha\Big\}
=\bigcup_{N\ge0}\Big\{\bm{x}\in\mathbb{R}^n \mid V_N(\bm{x})<\alpha\Big\},
\]
which is open as a union of open sets, since upper semicontinuity of $V_N$ is equivalent to openness of all its strict sublevel sets. Hence $V$ is upper semicontinuous on $\mathbb{R}^n$, and therefore on the compact set $\mathbb{X}$. \qed
\end{proof}

\begin{theorem}[Equivalence Between Robust Invariance and Positive Safety Probability]
Assume that the conditions of Lemma~\ref{thm:aps} hold, and that the support of $\mathrm{P}_{\bm{d}}$ is exactly $\mathbb{D}$, i.e., for every $\bm{d}\in\mathbb{D}$ and every $r>0$, $\mathrm{P}_{\bm{d}}[\mathbb{B}_r(\bm{d})]>0$, where $\mathbb{B}_r(\bm{d})$ denotes the open ball of radius $r$. Then the following statements are equivalent.
\begin{enumerate}
    \item[(1)] There exists a nonempty compact robust invariant set $\mathbb{S}\subseteq\mathbb{X}$, i.e.,
    \[\bm{f}(\bm{x},\bm{d})\in\mathbb{S},\qquad \forall(\bm{x},\bm{d})\in\mathbb{S}\times\mathbb{D},\]
    or equivalently,
    \[\bm{\phi}_{\pi}^{\bm{x}}(k)\in\mathbb{S},\qquad \forall(\bm{x},\pi,k)\in\mathbb{S}\times\mathbb{D}^{\infty}\times\mathbb{N}.\]
    \item[(2)] There exists a state $\bm{x}\in\mathbb{X}$ satisfying
    \[\mathrm{P}_{\pi}\Big[\bigwedge_{k=0}^{\infty}\bm{\phi}_{\pi}^{\bm{x}}(k)\in\mathbb{X}\Big]>0.\]
\end{enumerate}
\end{theorem}
\begin{proof}
$(1)\Rightarrow(2)$: take $\bm{x}\in\mathbb{S}$. Since $\bm{f}(\bm{x}',\bm{d})\in\mathbb{S}\subseteq\mathbb{X}$ for all $(\bm{x}',\bm{d})\in\mathbb{S}\times\mathbb{D}$, every trajectory starting at $\bm{x}$ remains in $\mathbb{X}$, so the probability in (2) equals $1>0$.

$(2)\Rightarrow(1)$: by Lemma~\ref{thm:aps}, there exists a nonempty set $\mathbb{S}\subseteq\mathbb{X}$ with
\begin{equation}\label{eq:proof1}
\mathrm{P}_{\pi}\Big[\bigwedge_{j=0}^{\infty}\bm{\phi}^{\bm{x}}_{\pi}(j)\in\mathbb{S}\Big]=1,\qquad \forall\bm{x}\in\mathbb{S}.
\end{equation}
Since $\mathbb{S}\subseteq\mathbb{X}$, \eqref{eq:proof1} implies that $V(\bm{x})=1$ for every $\bm{x}\in\mathbb{S}$, and hence $\max_{\bm{x}\in\mathbb{X}}V(\bm{x})=1$. We take $\mathbb{S}$ to be the set constructed in the proof of Lemma \ref{thm:aps}, namely $\mathbb{S}=\{\bm{x}\in\mathbb{X}\mid V(\bm{x})=1\}$. Since $V\le1$ everywhere, $\mathbb{S}=V^{-1}([1,\infty))$, which is closed because $V$ is upper semicontinuous (Lemma~\ref{upper_con}) and $[1,\infty)$ is closed. As $\mathbb{S}\subseteq\mathbb{X}$ and $\mathbb{X}$ is compact, $\mathbb{S}$ is compact.

We show that $\mathbb{S}$ is robustly invariant. The proof is inspired by the proof of Theorem 1 in \cite{wu2025controlled}. Suppose not: then there exist $\bm{x}_0\in\mathbb{S}$ and $\bm{d}_0\in\mathbb{D}$ with $\bm{f}(\bm{x}_0,\bm{d}_0)\notin\mathbb{S}$. Since $\mathbb{S}$ is closed, $\mathbb{R}^n\setminus\mathbb{S}$ is open, so there is $\delta>0$ with
\[
\mathbb{B}^{x}_{\delta}\big(\bm{f}(\bm{x}_0,\bm{d}_0)\big)\subseteq\mathbb{R}^n\setminus\mathbb{S},
\qquad
\mathbb{B}^{x}_{\delta}\big(\bm{f}(\bm{x}_0,\bm{d}_0)\big):=\big\{\bm{y} \mid \|\bm{y}-\bm{f}(\bm{x}_0,\bm{d}_0)\|<\delta\big\}.
\]
By continuity of $\bm{f}$ in $\bm{d}$ at $(\bm{x}_0,\bm{d}_0)$, there is $r>0$ such that
\[
\bm{f}(\bm{x}_0,\bm{d})\in\mathbb{B}^{x}_{\delta}\big(\bm{f}(\bm{x}_0,\bm{d}_0)\big)\subseteq\mathbb{R}^n\setminus\mathbb{S},
\qquad \forall\bm{d}\in\mathbb{B}^{d}_{r}(\bm{d}_0)\cap\mathbb{D},
\]
where $\mathbb{B}^{d}_{r}(\bm{d}_0):=\{\bm{d}\mid \|\bm{d}-\bm{d}_0\|<r\}$. Since $\mathrm{supp}(\mathrm{P}_{\bm{d}})=\mathbb{D}$, the relatively open neighborhood $\mathbb{B}^{d}_{r}(\bm{d}_0)\cap\mathbb{D}$ of $\bm{d}_0$ has positive probability, so
\[
\mathrm{P}_{\bm{d}}\big[\bm{f}(\bm{x}_0,\bm{d})\notin\mathbb{S}\big]
\ge\mathrm{P}_{\bm{d}}\big[\mathbb{B}^{d}_{r}(\bm{d}_0)\cap\mathbb{D}\big]>0.
\]
On the other hand, \eqref{eq:proof1} with $\bm{x}=\bm{x}_0$ and $j=1$, together with $\bm{\phi}^{\bm{x}_0}_{\pi}(1)=\bm{f}(\bm{x}_0,\bm{d}(0))$ and $\bm{d}(0)\sim\mathrm{P}_{\bm{d}}$, gives
\[
\mathrm{P}_{\bm{d}}\big[\bm{f}(\bm{x}_0,\bm{d})\in\mathbb{S}\big]
=\mathrm{P}_{\pi}\big[\bm{\phi}^{\bm{x}_0}_{\pi}(1)\in\mathbb{S}\big]=1,
\]
a contradiction. Hence $\bm{f}(\bm{x},\bm{d})\in\mathbb{S}$ for all $(\bm{x},\bm{d})\in\mathbb{S}\times\mathbb{D}$, i.e., $\mathbb{S}$ is robustly invariant. This completes the proof. \qed
\end{proof}

\setcounter{lemma}{2}
\begin{lemma}[State-uniform PAC Finite-time Reach-avoid Guarantee]
    Let $\epsilon, \delta  \in (0,1)$ and $\gamma \in (1,\infty)$ be given constants.
Let $\bm{a}^*(D^{M})$ denote the optimal solution to \eqref{eq:RBF_linear21}, selected according to the tie-break rule as suggested in  \cite{calafiore2006scenario} when the program \eqref{eq:RBF_linear21} admits multiple minimizers.  Define the certification procedure $\mathcal{A}(D^M)$ as in \eqref{cp}. Then, under Assumption \ref{as:h}, the probabilistic guarantee holds for every $\beta>0$:
 \begin{equation*}
\begin{split}
   &\textnormal{P}_{\bm{d}}^M \left[
    {\mathbb{X}_{s,\beta}\neq \emptyset} \Rightarrow
    \begin{split}
         \left(\forall \bm{x}\in \mathbb{X}_{s,\beta}\setminus \mathbb{S}, \textnormal{P}_{\pi}\left[
         \begin{split}
         &\exists k \leq K. \bm{\phi}_{\pi}^{\bm{x}} (k)\in \mathbb{S}~\wedge\\
         &\bigwedge_{j=0}^{k-1}\bm{\phi}_{\pi}^{\bm{x}}(j) \in \mathbb{X}_{s,\beta}
         \end{split}
         \right]\geq (1-\epsilon)^{K}\right)
        \end{split}
        \right]\\
        &\geq 
          1-\delta,
\end{split}
\end{equation*}
where $\mathbb{X}_{s,\beta}=\{\bm{x}\in \mathbb{X}\mid h(\bm{a}^*(D^M),\bm{x}) \geq \beta\}$,  $\sup_{\bm{x}\in \mathbb{X}\setminus\mathbb{S}} h(\bm{a}^*(D^M),\bm{x}) \leq H$, $K=\lfloor{\log_{\gamma}\frac{H}{\beta}\rfloor}+1$, and the sample size $M$ satisfies \eqref{MM_so}.
\end{lemma}
\begin{proof}
By Lemma \ref{lemma:robust_s}, with confidence at least $1-\delta$ over the sampled disturbance set $D^M$, if $\mathbb{X}_{s,\beta}\neq \emptyset$, then for every $\bm{x}\in \mathbb{X}_{s,\beta}\setminus \mathbb{S}$,
\[\textnormal{P}_{\bm{d}}\left[
    h(\bm{a}^*(D^M),\bm{f}(\bm{x},\bm{d})) \geq \gamma h(\bm{a}^*(D^M),\bm{x})  \bigwedge  \bm{f}(\bm{x},\bm{d}) \in \mathbb{X}_s
     \right]\geq 1-\epsilon.\]

We define the first hitting time (stopping time) to the target set $\mathbb{S}$ as:
\[\tau:=\inf\{k\geq 0\mid \bm{\phi}_{\pi}^{\bm{x}}(k)\in \mathbb{S}\}.\]

     Define the successful one-step transition event
     \[E_i:=\{\tau\leq i-1\}\bigcup \left\{\begin{split}
     &\tau>i-1, \bm{\phi}_{\pi}^{\bm{x}}(i)\in \mathbb{X}_{s,\beta},\\
     &h(\bm{a}^*(D^M),\bm{\phi}_{\pi}^{\bm{x}}(i))\geq \gamma h(\bm{a}^*(D^M),\bm{\phi}_{\pi}^{\bm{x}}(i-1))
     \end{split}\right\} \]
     \color{black}
     for $i=1,\ldots,K$.

Let $\mathcal{F}_{i-1}$ be the $\sigma$-algebra generated by the trajectory up to time $i-1$. We evaluate the conditional probability of $E_i$ given the past trajectory. 

Because $\{\tau\leq i-1\} \in \mathcal{F}_{i-1}$, we can partition the conditioning based on whether the target has been reached. 
\begin{enumerate}
    \item On the event $\{\tau\leq i-1\}$: Here $E_i$ holds by the first branch of its definition, and $\{\tau\le i-1\}\in\mathcal{F}_{i-1}$ because $\tau$ is a stopping time with respect to $(\mathcal{F}_{j})_{j\ge0}$. Since $1_{E_i}\ge 1_{\{\tau\le i-1\}}$ and $1_{\{\tau\le i-1\}}$ is $\mathcal{F}_{i-1}$-measurable,
\[
\mathrm{E}_{\pi}\big[1_{E_i}\mid\mathcal{F}_{i-1}\big]\ \ge\ 1_{\{\tau\le i-1\}} ,
\]
so $\mathrm{P}_{\pi}[E_i\mid\mathcal{F}_{i-1}]=1$ $\mathrm{P}_{\pi}$-almost surely on $\{\tau\le i-1\}$.
\item On the event $\{\tau>i-1\} \cap \bigcap_{j=1}^{i-1} E_j$: the target has not been reached, and all previous steps were successful. This implies the current state $\bm{\phi}_{\pi}^{\bm{x}}(i-1)\in \mathbb{X}_{s,\beta}\setminus \mathbb{S}$. Because the system is Markovian and the disturbances are i.i.d., the future evolution depends only on $\bm{\phi}_{\pi}^{\bm{x}}(i-1)$. By the one-step PAC guarantee established in Lemma \ref{lemma:robust_s}, 
\[
\begin{split}
&\textnormal{P}_{\pi}\left[E_i\middle | \mathcal{F}_{i-1}\right]\\
=&\textnormal   {P}_{\bm{d}}\left[h(\bm{a}^*(D^M),\bm{\phi}_{\pi}^{\bm{x}}(i)) \geq \gamma h(\bm{a}^*(D^M),\bm{\phi}_{\pi}^{\bm{x}}(i-1))  \bigwedge  \bm{\phi}_{\pi}^{\bm{x}}(i) \in \mathbb{X}_{s,\beta}\mid \bm{\phi}_{\pi}^{\bm{x}}(i-1)\right]\\
\geq &1-\epsilon.
\end{split}
\]
\end{enumerate}

 Fix any $i\geq 2$. Consequently, it holds almost surely on $\bigcap_{j=1}^{i-1}E_j$ that 

\[
\begin{split}
\textnormal{P}_{\pi}\left[E_i\middle | \bigcap_{j=1}^{i-1} E_j\right]=&\textnormal{E}_{\pi}\left[1_{E_i}\middle |\bigcap_{j=1}^{i-1} E_j\right]\\
=&\textnormal{E}_{\pi}\left[\textnormal{E}_{\pi}\left[1_{E_i}\middle | \mathcal{F}_{i-1} \right]\middle |\bigcap_{j=1}^{i-1} E_j\right]
=\textnormal{E}_{\pi}\left[\textnormal{P}_{\pi}\left[E_i\middle | \mathcal{F}_{i-1}  \right]\middle |\bigcap_{j=1}^{i-1} E_j\right]\\
\geq &1-\epsilon.
\end{split}
\]

     Applying the chain rule for conditional probabilities yields 
     \[\textnormal{P}_{\pi}\left[ \bigcap_{i=1}^K E_i\right]=\prod_{i=1}^K \textnormal{P}_{\pi}\left[E_i\middle | \bigcap_{j=1}^{i-1} E_j\right]\geq (1-\epsilon)^K.\]

Finally, since $H$ is an upper bound of $h$ on $\mathbb{X}\setminus\mathbb{S}$, if $\tau > K$ and $\bm{\phi}_{\pi}^{\bm{x}}(i)\in \mathbb{X}\setminus\mathbb{S}, \forall i =0,\ldots,\tau$, 
$h(\bm{\phi}_{\pi}^{\bm{x}}(\tau))> h(\bm{\phi}_{\pi}^{\bm{x}}(K))\geq \gamma^K h(\bm{x})\geq \gamma^K \beta > H$ is a contradiction. Thus, on the event $\bigcap_{i=1}^K E_i$, the stopping time satisfies $\tau\leq K$.

Because $\bigcap_{i=1}^K E_i$ implies both that the target set is reached within $K$ steps ($\exists k \leq K. \bm{\phi}_{\pi}^{\bm{x}} (k)\in \mathbb{S}$) and that all states prior to reaching $\mathbb{S}$ remained in $\mathbb{X}_{s,\beta}$, it follows that 
\[
\textnormal{P}_{\pi}\left[\exists k \leq K. \bm{\phi}_{\pi}^{\bm{x}} (k)\in \mathbb{S}~\wedge \bigwedge_{j=0}^{k-1}  \bm{\phi}_{\pi}^{\bm{x}}(j) \in \mathbb{X}_{s,\beta}\right]\geq \textnormal{P}_{\pi}\left[\bigcap_{i=1}^K E_i\right] \geq  (1-\epsilon)^{K}.
\]

 Since the one-step guarantee holds with confidence at least $1-\delta$ over the sampled dataset, the same confidence level applies to the above finite-time reach-avoid guarantee. This completes the proof.     
    \qed
\end{proof}

\setcounter{theorem}{3}
\begin{theorem}[PAC Guarantee of the SOS Implementation]
\label{cor:sos_pac}
Let $\epsilon,\delta\in(0,1)$ and $\gamma>1$, and let $\widetilde D^{M_1}$ and $D^M$ be mutually independent datasets of i.i.d. disturbance samples drawn from $\mathbb P_{\bm d}$. Denote the optimal solution to \eqref{eq:rbc2_sos} by $\bm a_{\rm sos}^*(\widetilde D^{M_1},D^M)$. Under Assumption~\ref{as:h}, if
\begin{equation*}
M\geq\frac{2}{\epsilon}
\left(\ln\frac{1}{\delta}+m+1\right),
\end{equation*}
then the PAC infinite-time safety guarantee in Theorem~\ref{thm:so} holds for $h(\bm a_{\rm sos}^*,\cdot)$. The sets and quantities $\mathbb X_{s,\beta}$, $H$, and $K$ are defined using this barrier function as in Theorem~\ref{thm:so}.
\end{theorem}

\begin{proof}
We first show that the SOS multipliers $s_3$ and $s_4$ do not increase the required sample size. Define
\[
\mathcal C_{\bm a}= \left\{\bm a\in[-U_a,U_a]^m\ \middle|
\begin{aligned}
&\exists\,\bm c_3\in[-U_s,U_s]^{q_3},\quad
\bm c_4\in[-U_s,U_s]^{q_4}: s_3,s_4\in\Sigma[\bm x],\\
&-h(\bm a,\bm x)+s_3(\bm x)\hat g(\bm x)
-s_4(\bm x)g(\bm x)\in\Sigma[\bm x]
\end{aligned}
\right\}.
\]
Thus, $\mathcal C_{\bm a}$ consists precisely of the barrier parameters for which admissible multipliers satisfying the deterministic SOS constraint exist.

Under Assumption~\ref{as:h}, these SOS constraints are affine in $(\bm a,\bm c_3,\bm c_4)$. Together with the coefficient bounds, they define a compact convex set, since the fixed-degree SOS cones are closed and convex. The set $\mathcal C_{\bm a}$, obtained by projecting this joint feasible set onto $\bm a$, is therefore compact and convex and is independent of $D^M$.

Neither the objective nor the sampled constraints in \eqref{eq:rbc2_sos} depend on $\bm c_3$ or $\bm c_4$. Consequently, eliminating these coefficients and requiring
\[
(\bm a,\kappa)\in \mathcal C_{\bm a}\times[0,\bar\kappa] \subseteq\mathbb R^{m+1}
\]
preserves the objective and all feasible parameter pairs of \eqref{eq:rbc2_sos}. The multipliers $s_3$ and $s_4$ therefore affect only the deterministic constraint set, while the resulting scenario program involves $m+1$ parameters. 

Since the SOS constraints are sufficient for the corresponding barrier conditions, this scenario guarantee yields the one-step PAC conclusion of Lemma~\ref{lemma:robust_s}. The conclusions of Lemma~\ref{lemma:so} and Theorem~\ref{thm:so} then follow by the same arguments as in the main text. \qed
\end{proof}

\section{Details of the SOS Programming}
\label{sec:app_sos}

This section provides the SOS formulations used in the numerical experiments.

Given the computed target set $\mathbb{S}$, the uncertain optimization problem \eqref{eq:RBF_linear1} is relaxed as
\begin{align}
\label{eq:rbc1_sos}
    &\min_{\bm{a},s_1,s_2,s_3,s_4,s_5}
    \quad-\frac{1}{N_o}\sum_{i=1}^{N_o}
    h(\bm{a},\bm{x}'_i)\\
    \text{s.t.}\quad
    &\begin{cases}
    h(\bm{a},\bm{f}(\bm{x},\bm{d}))
    -\gamma h(\bm{a},\bm{x})
    +s_1(\bm{x},\bm{d})g(\bm{x})\\
    \hspace{2cm}+s_2(\bm{x},\bm{d})v(\bm{x})
    +s_3(\bm{x},\bm{d})p(\bm{d})
    \in\Sigma[\bm{x},\bm{d}],\\
    -h(\bm{a},\bm{x})
    +s_4(\bm{x})\hat g(\bm{x})
    -s_5(\bm{x})g(\bm{x})
    \in\Sigma[\bm{x}],\\
    s_1(\bm{x}, \bm{d}),s_2(\bm{x},\bm{d}),s_3(\bm{x},\bm{d}) \in \Sigma [\bm{x},\bm{d}]; ~s_4(\bm{x}),s_5(\bm{x}) \in \Sigma [\bm{x}],\\
    \bm{a}\in[-U_a,U_a]^{m}.
    \end{cases}
    \notag
\end{align}

Besides, after obtaining $h(\bm{x})$, the upper bound $H$ required in Theorem~\ref{thm:so} is computed by
\begin{align}
\label{eq:H_sos}
    &\min_{H,s_1,s_2} \quad H\\
    \text{s.t.}&\begin{cases}
     H - h(\bm{a},\bm{x}) + s_1(\bm{x}) g(\bm{x}) + s_2(\bm{x}) v(\bm{x}) \in \Sigma[\bm{x}], \\
        s_1(\bm{x}),s_2(\bm{x}) \in \Sigma[\bm{x}].\\
    \end{cases}\notag
\end{align}

\section{Experimental setup}
\label{sec:app_exp}

This section provides additional details of the experimental setup. We first specify the dynamical systems and disturbance distributions used in the experiments summarized in Table~\ref{tab:exp}, and then describe the parameter settings and random seeds used throughout the experiments.

All computations are run on a machine equipped with an i7-13700 2.1GHz CPU with 32GB RAM. The resulting semi-definite programs are implemented in YALMIP~\cite{lofberg2004yalmip} and solved using MOSEK~\cite{aps2019mosek}.

\begin{example}
\label{ex:3}
Consider the following two-dimensional nonlinear system adapted from \cite{zhang2018safety}:
\begin{equation*}
\begin{cases}
\bm{x}_1(k+1)=\bm{x}_1(k)+\tau_s\big(-\bm{x}_1(k)+2\bm{x}_1^2(k)\bm{x}_2(k)+\bm{d}_1(k)\bigl(0.015+\bm{x}_1(k)\bigr)\big),\\
\bm{x}_2(k+1)=\bm{x}_2(k)+\tau_s\big(-\bm{x}_2(k)+\bm{d}_2(k)\bigl(0.015+\bm{x}_2(k)\bigr)\big),
\end{cases}
\end{equation*}
where $\tau_s=0.9$. The disturbance vector $\bm d(k)=(\bm{d}_1(k),\bm{d}_2(k))^\top$ has independent components, each following a zero-mean normal distribution with standard deviation $0.03$, truncated to $[-0.78,0.78]$. The safe set is 
$\mathbb X = \left\{ \bm x\in\mathbb R^2\,\middle|\, \bm{x}_1^2+\bm{x}_2^2\leq 1\right\}$.
\end{example}

\begin{example}
\label{ex:4}
Consider the following nonlinear oscillator adapted from \cite{wang2023safety}:
\begin{equation*}
\begin{cases}
\begin{split}
&\bm{x}_1(k+1)=\bm{x}_1(k)+\tau_s \bm{x}_2(k),\\
&\bm{x}_2(k+1)=\bm{x}_2(k)+\tau_s \Big(-\bm{x}_1(k)+\dfrac{1}{3}\bm{x}_1^3(k)-\bm{x}_2(k)\\
&\hspace{3cm}+\bigl(2.5+1.5d(k)\bigr)\bigl(1.04-\bm{x}_1(k)-\bm{x}_2(k)\bigr)\Big),
\end{split}
\end{cases}
\end{equation*}
where $\tau_s=0.3$. The safe set
is $\mathbb X=\left\{\bm x\in\mathbb R^2\,\middle|\,\bm{x}_1^2+\bm{x}_2^2\leq 1\right\}$. The disturbance $d(k)$ follows a scaled Beta distribution on $[-1,1]$, obtained from a $\operatorname{Beta}(100,800)$ distribution. 
\end{example}

\begin{example}
\label{ex:5}
Consider the discretization of the Moore-Greitzer model of a jet engine:
\begin{equation*}
\begin{cases}
\bm{x}_1(k+1)=\bm{x}_1(k)+\tau_s\big(-\bm{x}_2(k)-1.5\bm{x}_1^2(k)-0.5\bm{x}_1^3(k)+d(k)\bm{x}_1(k)\big),\\
\bm{x}_2(k+1)=\bm{x}_2(k)+\tau_s\big(3\bm{x}_1(k)-\bm{x}_2(k)+d(k)\bm{x}_2(k)\big),
\end{cases}
\end{equation*}
where $\tau_s=0.25$. The disturbance $d(k)$ follows a truncated normal distribution, defined on $[-1.72,-0.22]$ with a mean of $-1.5$ and a standard deviation of $0.01$. The safe set is given by $\mathbb X=\left\{\bm x\in\mathbb R^2\,\middle|\,\bm{x}_1^2+\bm{x}_2^2\leq 1\right\}$. 
\end{example}

\begin{example}
\label{ex:6}
Consider the following two-dimensional uncertain polynomial system adapted from \cite{sogokon2016non}:
\begin{equation*}
\begin{cases}
\bm{x}_1(k+1)=\bm{x}_1(k)+\tau_s\big(-2\bm{x}_1(k)+\bm{x}_1^2(k)+\bm{x}_2(k)+d(k)\bm{x}_2(k)\big),\\
\bm{x}_2(k+1)=\bm{x}_2(k)+\tau_s\big(\bm{x}_1(k)-2\bm{x}_2(k)+\bm{x}_2^2(k)-d(k)\bm{x}_1(k)\big),
\end{cases}
\end{equation*}
where $\tau_s=0.5$. The disturbance $d(k)$ follows a zero-mean normal distribution with standard deviation $0.02$, truncated to $[-0.82,0.82]$. The safe set is $\mathbb X=\left\{\bm x\in\mathbb R^2\,\middle|\,\bm{x}_1^2+\bm{x}_2^2\leq 1\right\}$. 
\end{example}

\begin{example}
\label{ex:rotor3d}
Consider a stochastic damped rigid-body rotor based on the Euler angular-momentum equations, adapted from \cite{hanna2021integrable}. Its forward Euler discretization is given by
\begin{equation*}
\begin{cases}
x_1(k+1)=x_1(k)+\tau_s\big(
-x_1(k)-0.5x_2(k)x_3(k)-d(k)x_2(k)
\big),\\
x_2(k+1)=x_2(k)+\tau_s\big(
-x_2(k)+x_1(k)x_3(k)+d(k)x_1(k)
\big),\\
x_3(k+1)=x_3(k)+\tau_s\big(
-x_3(k)-0.5x_1(k)x_2(k)
\big),
\end{cases}
\end{equation*}
where $\tau_s=0.7$. The deterministic quadratic terms describe the exchange of angular momentum among the three principal axes, while $d(k)$ represents a random transverse torque. The disturbance $d(k)$ follows a truncated normal distribution supported on $[-1.19,1.19]$, with mean $0$ and standard deviation $0.04$. The safe set is the three-dimensional unit ball $\mathbb X= \left\{\bm x\in\mathbb R^3\,\middle|\,x_1^2+x_2^2+x_3^2\leq 1\right\}$.
\end{example}

\begin{example}
\label{ex:shg4d}
Consider a stochastic quadratic coupled-mode model for second-harmonic generation in a doubly resonant nonlinear cavity, adapted from \cite{rodriguez2007chi}. Let
\[
A=x_1+\mathrm{i}x_2,
\qquad
B=x_3+\mathrm{i}x_4
\]
denote the complex amplitudes of the fundamental and second-harmonic modes, respectively. After normalization and forward Euler discretization, the resulting four-dimensional system is
\begin{equation*}
\begin{cases}
x_1(k+1)=x_1(k)+\tau_s\big(
-x_1(k)+x_1(k)x_3(k)+x_2(k)x_4(k)-d(k)x_2(k)
\big),\\
x_2(k+1)=x_2(k)+\tau_s\big(
-x_2(k)+x_1(k)x_4(k)-x_2(k)x_3(k)+d(k)x_1(k)
\big),\\
x_3(k+1)=x_3(k)+\tau_s\big(
-x_3(k)-x_1^2(k)+x_2^2(k)-2d(k)x_4(k)
\big),\\
x_4(k+1)=x_4(k)+\tau_s\big(
-x_4(k)-2x_1(k)x_2(k)+2d(k)x_3(k)
\big),
\end{cases}
\end{equation*}
where $\tau_s=0.7$. The quadratic terms describe energy exchange between the fundamental and second-harmonic modes. The disturbance $d(k)$ models a random normalized detuning and follows a truncated normal distribution supported on $[-0.45,0.45]$, with mean $0$ and standard deviation $0.04$. The safe set is $\mathbb X=\left\{\bm x\in\mathbb R^4\,\middle|\,x_1^2+x_2^2+x_3^2+x_4^2\leq 1\right\}$.
\end{example}

\begin{example}
\label{ex:6d}
We consider the following six-dimensional uncertain
polynomial system:
\begin{equation*}
x_i(k+1)
=
0.5x_i(k)+1.2d(k)x_i^2(k),
\qquad i=1,\ldots,6.
\end{equation*}
The scalar disturbance $d(k)$ follows a zero-mean normal distribution with standard deviation $0.02$, truncated to $[-1,1]$. The safe set is the
six-dimensional unit ball $\mathbb X=\left\{\bm x\in\mathbb R^6\,\middle|\,\|\bm x\|_2^2\leq 1\right\}$.
\end{example}

For Examples \ref{ex:1} and \ref{ex:6d}, the maximal robust invariant sets can be characterized directly from the system dynamics. In Example \ref{ex:1}, letting $r^2(k)=x^2(k)+y^2(k)$ gives $r(k+1)=|0.5+(0.4+r^2(k))d(k)|~r(k)$, from which robust invariance holds exactly for $r^2(k)\leq 0.1$. In Example \ref{ex:6d}, $|\bm x_i(k+1)|\leq 0.5|\bm x_i(k)|+1.2|\bm x_i(k)|^2\leq |\bm x_i(k)|$ for all $d(k)\in[-1,1]$ whenever $|\bm x_i(k)|\leq 5/12$, whereas an admissible disturbance can make any coordinate with $|\bm x_i(k)|>5/12$ grow. Hence, the maximal robust invariant set is $\mathbb X\cap[-5/12,5/12]^6$. These analytic characterizations provide exact robust-invariance references in Table~\ref{tab:exp}. For the reach-avoid SOS computation in Example \ref{ex:6d}, we use the simpler robust invariant subset $\mathbb S=\{\bm x\in\mathbb X\mid \|\bm x\|_2^2\leq(5/12)^2\}$ as the target set, which preserves the single-polynomial target representation used in Appendix~\ref{sec:app_sos}; the ``Robust Invariant Set'' column of Table~\ref{tab:exp} reports the maximal set described above.

We next provide details of the Monte Carlo simulations used in Example~\ref{ex:lotka} for empirical validation of the proposed certification. For the one-step validation, we uniformly sample $10^5$ states from the certified region outside the robust invariant target set and, for each sampled state, draw $10^3$ independent disturbances. The empirical one-step satisfaction probability is estimated as the fraction of successor states that remain in the certified region and satisfy
\[
h(\bm a_{\rm sos}^*(\widetilde D^{M_1},D^M),\bm{f}(\bm{x},\bm{d})) \geq \gamma h(\bm a_{\rm sos}^*(\widetilde D^{M_1},D^M),\bm{x}).
\]
For the long-horizon validation, we uniformly sample $10^5$ initial states from the same region. For each initial state, we simulate $10^3$ independent trajectories, each over $10^3$ steps. The empirical safety rate is estimated as the fraction of simulated trajectories from that initial state that remain in the safe set $\mathbb{X}$ throughout the entire simulation horizon.

\paragraph{Parameter settings.}
We next summarize the parameter settings used throughout the experiments. For consistency in the numerical comparison, the sample-based RBGF program of the finite-time PAC baseline~\cite{wu2026pac} is implemented using the same two-stage SOS construction as the proposed method, while its original barrier condition and PAC guarantee are kept unchanged. The coefficient bounds for the barrier functions and SOS multipliers are fixed as $U_{\bm a}=U_{\bm s}=10$ in all cases. In all experiments, we set the size of the auxiliary dataset to be the same as that of the scenario dataset $D^M$, i.e., $M_1=M$. For all methods, the number of state samples used in the optimization objectives is fixed at $1000$, i.e., $N_r=N_o=1000$. For the finite-time PAC baseline~\cite{wu2026pac}, we follow the setting used in the original work and fix $\gamma=0.99$ in all experiments. 

For a fair comparison, within each example we use the same polynomial degree for all corresponding barrier function and SOS multiplier templates. Table~\ref{tab:para} summarizes the polynomial degree, the sample sizes $M$ used by the proposed method and the finite-time PAC baseline~\cite{wu2026pac}, the value of $\gamma$ used in the robust invariant-set computation~\eqref{eq:target_sos}, the value of $\gamma$ used by the proposed RBGF formulation~\eqref{eq:rbc_aux_sos} and \eqref{eq:rbc2_sos}, and the value of $\gamma$ used in the worst-case RBGF SOS program~\eqref{eq:rbc1_sos}. For the proposed method, $\gamma$ is selected using only the auxiliary dataset $\widetilde D^{M_1}$ and fixed before the independent certification dataset $D^M$ is drawn.

\begin{table*}[!htb]
\centering
\caption{Parameter settings used in the experiments.}
\setlength{\tabcolsep}{3mm}{
\begin{tabular}{ccccccccccc}
\hline
Ex & Polynomial Degree & M & $\gamma$ in \eqref{eq:target_sos} & $\gamma$ in \eqref{eq:rbc1_sos} & $\gamma$ in \eqref{eq:rbc_aux_sos} and \eqref{eq:rbc2_sos} \\
\hline
\ref{ex:1}     & 4 & 1528 & -      & - & 1.1\\
\ref{ex:lotka} & 6 & 2394 & 0.9999 & 1.1   & 1.5\\
\ref{ex:3}     & 6 & 2394 & 0.99   & 1.01  & 1.3\\
\ref{ex:4}     & 6 & 2394 & 0.9    & 1.1  & 1.4\\
\ref{ex:5}     & 4 & 1528 & 0.99   & 1.01  & 1.65\\
\ref{ex:6}     & 6 & 2394 & 0.99   & 1.1  & 1.6\\
\ref{ex:rotor3d}&2 & 1194 & 1.0    & 1.1  & 1.8\\
\ref{ex:shg4d} & 2 & 1528 & 1.0    & 1.1  & 1.7\\
\ref{ex:6d}    & 2 & 2394 & -      & -  & 1.1\\
\hline
\end{tabular}}
\label{tab:para}
\end{table*}
For both the proposed method and the worst-case RBGF baseline, the reported certified volume is computed for the union of the corresponding RBGF-certified region and the robust invariant target set $\mathbb S$, since every state in $\mathbb S$ is already certified to remain safe for all future times by robust invariance.

\paragraph{Random seeds.}
To ensure reproducibility and fair comparisons, we use common random samples across different methods whenever applicable. We use a separate random seed for each of the following five sampling procedures:
(i) the state samples used in the objective of the robust invariant-set computation;
(ii) the state samples used to approximate the objective functions of the proposed method, the worst-case RBGF baseline, and the finite-time PAC baseline~\cite{wu2026pac};
(iii) the disturbance samples used in the auxiliary SOS program~\eqref{eq:rbc_aux_sos} of the proposed method;
(iv) the disturbance samples used to construct the scenario constraints of the proposed method and the finite-time PAC baseline; and
(v) the Monte Carlo samples used to estimate the relative volumes of the computed sets.
We consider three seed configurations,
\[
(0,1,2,3,4),\qquad
(100,101,102,103,104),\qquad
(200,201,202,203,204),
\]
where the five entries correspond, in order, to the five sampling procedures above. Examples~\ref{ex:1} and~\ref{ex:lotka} use the first configuration, whereas the results in Table~\ref{tab:exp} are reported over all three configurations.

\section{Additional Experimental Results}
\label{sec:app_ex_results}
In this section, we present additional experimental results. We examine the effects of $\epsilon$, the polynomial degree, and $\gamma$, illustrate how the proposed method uses disturbance samples to capture distributional information, and report results across different random-seed configurations.

\begin{figure}[!htbp]
  \centering
  \begin{minipage}[c]{0.45\linewidth}
    \centering
    \setlength{\tabcolsep}{2mm}
    \begin{tabular}{ccc}
      \hline
      $\epsilon$ & $M$ & Time (s) \\ \hline
      0.15 & 306 & 20.7\\
      0.10 & 459 & 40.9\\
      0.05 & 917 & 106.7\\
      0.03 &1528 & 179.1\\ \hline
    \end{tabular}
    \captionof{table}{Sample size and computation time for different $\epsilon$.}
    \label{tab:exp1_epsilon}
  \end{minipage}
  \hfill
  \begin{minipage}[c]{0.54\linewidth}
    \centering
    \includegraphics[width=\linewidth]{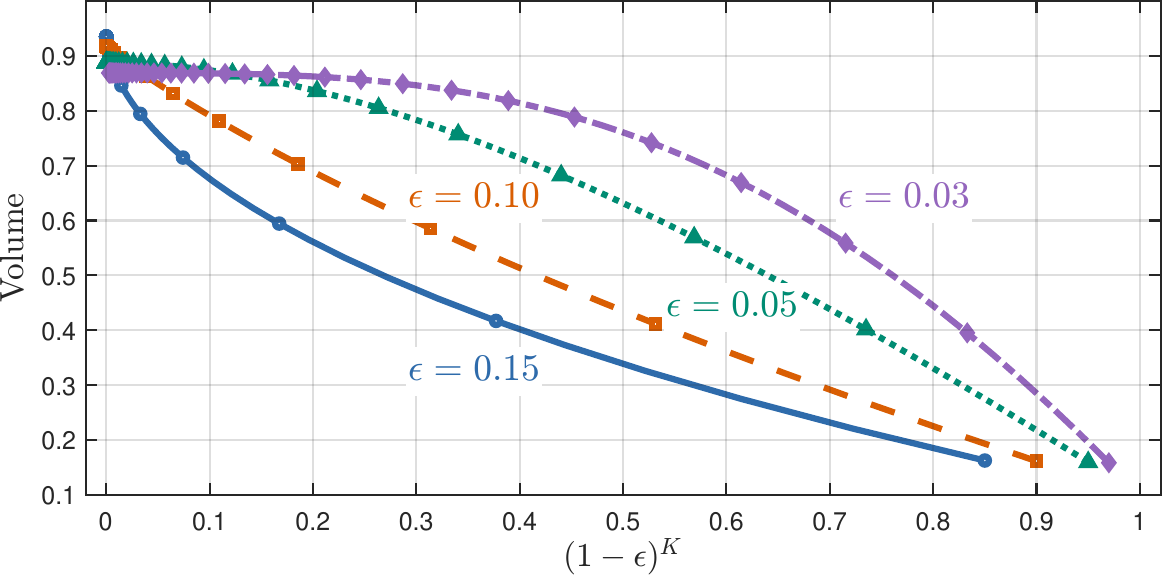}
    \captionof{figure}{Results for different $\epsilon$.}
    \label{fig:exp1_epsilon}
  \end{minipage}
\end{figure}

\paragraph{Effect of $\epsilon$.} We examine the effect of $\epsilon$ on certification in Example~\ref{ex:1}, with results summarized in Table~\ref{tab:exp1_epsilon} and Fig.~\ref{fig:exp1_epsilon}. For each fixed $\epsilon$, Fig.~\ref{fig:exp1_epsilon} plots the relative volume of $\mathbb{X}_{s,\beta}$ against the corresponding infinite-time safety probability lower bound $(1-\epsilon)^K$, with different points obtained by varying $\beta$. Decreasing $\epsilon$ tightens this lower bound, but it also increases the required sample size $M$ in \eqref{MM_so_sos}, raising the computational cost (Table~\ref{tab:exp1_epsilon}). Moreover, the additional samples enter \eqref{eq:rbc2_sos} as extra constraints, making the program more restrictive and thereby reducing the maximum certified volume, as visible in Fig.~\ref{fig:exp1_epsilon}. The parameter $\epsilon$ should therefore be chosen to balance guarantee strength, certified volume, and sample and computational cost.

\begin{figure}[!htbp]
  \centering
  \begin{minipage}[c]{0.4\linewidth}
    \centering
    \setlength{\tabcolsep}{2.0mm}
    \begin{tabular}{ccc}
    \hline
    Degree & $M$ & $\text{Time (s)}$  \\ \hline
    4  & 306 & 20.7\\
    6  & 479 & 151.9\\
    8  & 706 & 1028\\\hline
    \end{tabular}
    \captionof{table}{Sample size and computation time for different polynomial degrees.}
    \label{tab:exp1_degree}
  \end{minipage}
  \hfill
  \begin{minipage}[c]{0.55\linewidth}
    \centering
    \includegraphics[width=\linewidth]{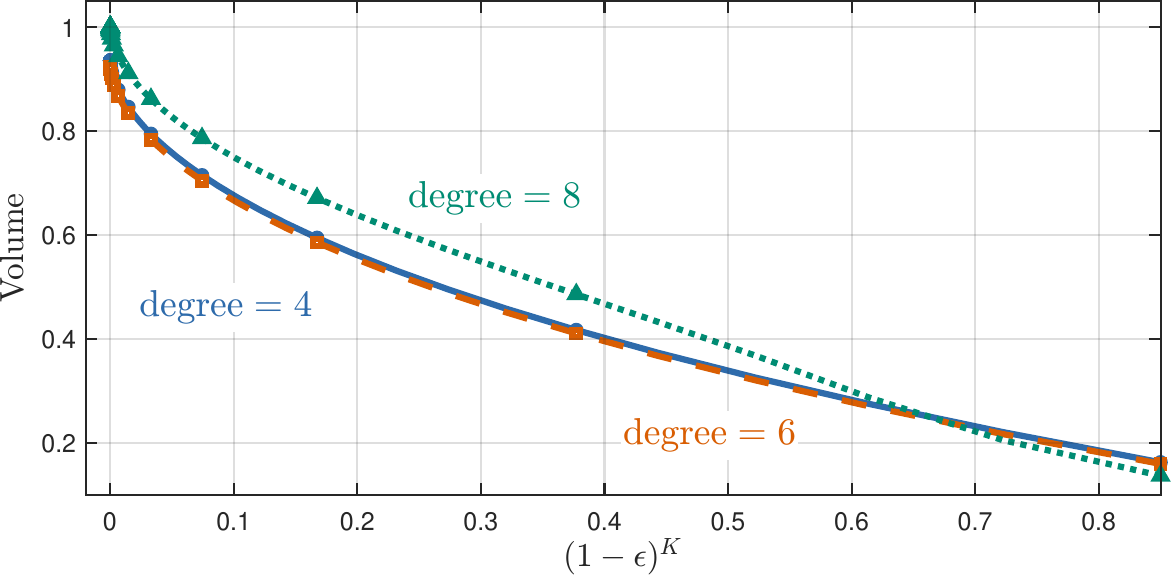}
    \captionof{figure}{Results for Example~\ref{ex:1} for different polynomial degrees.}
    \label{fig:exp1_degree}
  \end{minipage}
\end{figure}

\paragraph{Effect of the polynomial degree.} We next study the effect of the polynomial degree using Example~\ref{ex:1}, with $\epsilon$ fixed at $0.15$; Fig.~\ref{fig:exp1_degree} plots the relative volume of $\mathbb{X}_{s,\beta}$ against the lower bound $(1-\epsilon)^K$, with different points obtained by varying $\beta$. A higher degree increases the expressive power of $h(\bm{a},\bm{x})$ and can improve certification—the degree-$8$ template yields larger certified regions than degree $4$—but the improvement is not monotonic. A higher degree also increases the number of parameters $m$ and hence the required sample size $M$ in \eqref{MM_so_sos}, which raises the computational cost (Table~\ref{tab:exp1_degree}) and adds sampled constraints to \eqref{eq:rbc2_sos}, restricting the feasible set and offsetting the gain in expressiveness. The results for degrees $4$ and $6$ are nearly identical, illustrating these competing effects. The degree should therefore be chosen to balance expressiveness against sample and computational cost.

\begin{figure*}[!htbp]
    \centering
     \subfigure[Disturbance distributions.]{
    \includegraphics[height=3.5cm, width=4.5cm]{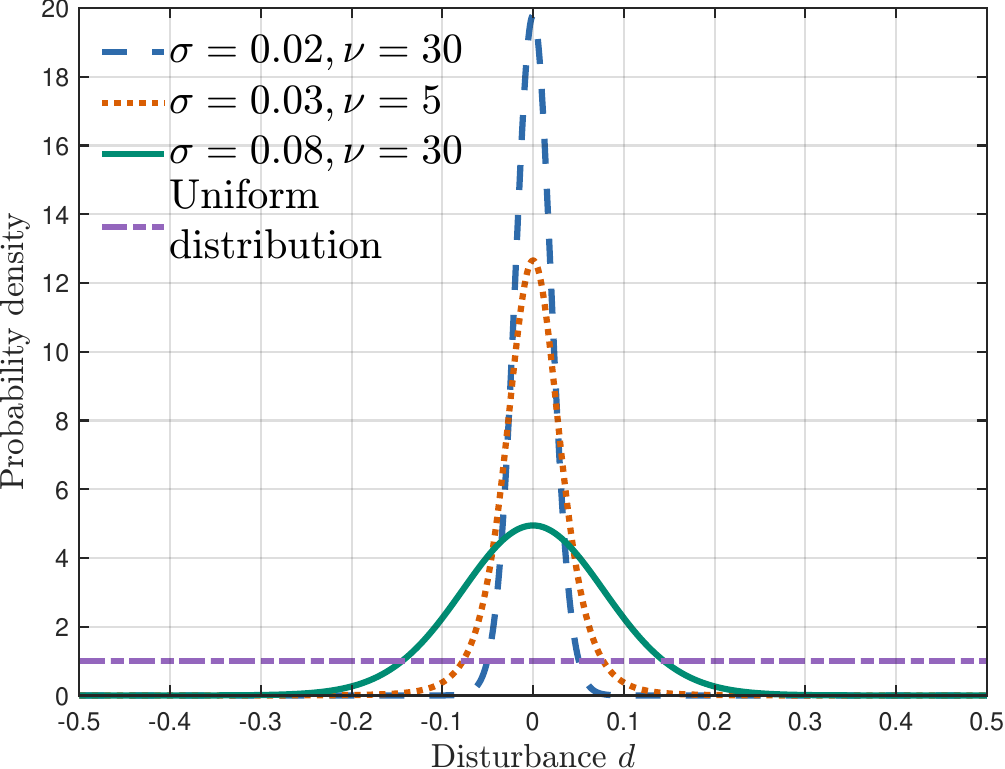}
    \label{fig:lotka_distribution1}
    }
    \subfigure[Results under different distributions.]{
    \includegraphics[height=3.5cm]{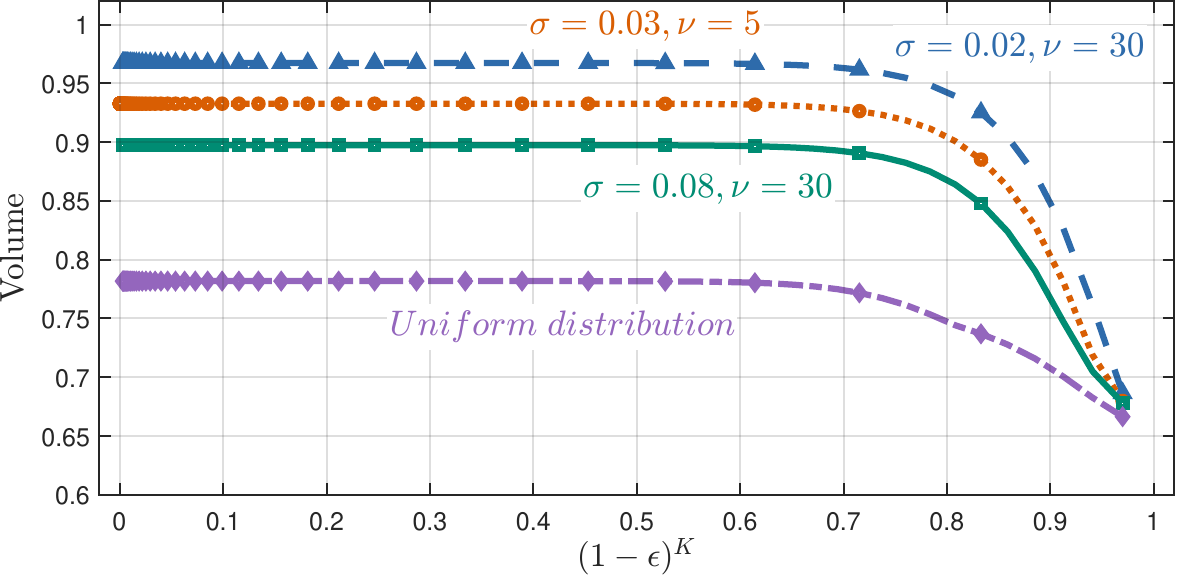}
    \label{fig:lotka_distribution2}
    }
    \caption{Results for Example~\ref{ex:lotka} under different disturbance distributions. $\sigma$ and $\nu$ denote the scale parameter and the number of degrees of freedom, respectively.}
    \label{fig:lotka_distribution}
\end{figure*}

\paragraph{Adaptation to Disturbance Distributions.} 
We use Example~\ref{ex:lotka} to examine how the proposed method adapts to different disturbance distributions and demonstrate how the proposed method exploits distributional information. We consider four disturbance distributions with the same support $\mathbb{D}$ but different densities: three truncated Student's $t$ distributions and a uniform distribution (Fig.~\ref{fig:lotka_distribution1}). Since the robust invariant-set computation and the worst-case RBGF formulation depend only on $\mathbb{D}$, they yield identical results across all four distributions. In contrast, the proposed method leverages finite disturbance samples to capture distributional characteristics while retaining the PAC confidence guarantee, without requiring an analytical form of $\mathrm{P}_{\bm{d}}$. As shown in Fig.~\ref{fig:lotka_distribution2}, distributions more concentrated around zero generally yield larger certified regions at a given infinite-horizon safety probability level. 

\begin{figure*}[!htbp]
    \centering
    \includegraphics[width=0.7\linewidth]{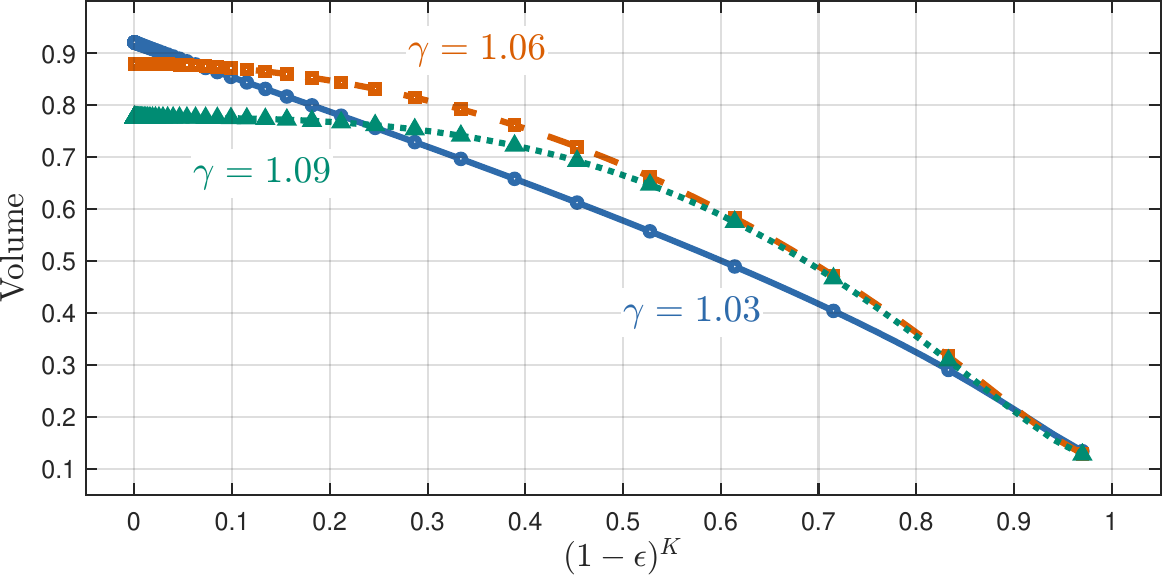}
    \caption{Results for Example~\ref{ex:lotka} under different values of $\gamma$.}
    \label{fig:lotka_gamma}
\end{figure*}

\paragraph{Effect of $\gamma$.} We further investigate the effect of the parameter $\gamma$ using Example~\ref{ex:lotka}. To make its influence more evident, we consider the setting with the smaller robust invariant set as the target set while keeping all other parameters fixed. The results under different values of $\gamma$ are shown in Fig.~\ref{fig:lotka_gamma}. As shown in Fig.~\ref{fig:lotka_gamma}, $\gamma$ dictates the tradeoff between the certified volume and the strength of the guarantee. However, the effect of increasing $\gamma$ is non-monotonic—the tradeoff curves cross—so there is no single choice of $\gamma$ that dominates the others. Determining an optimal $\gamma$ is system-dependent and beyond the scope of this work.

\paragraph{Variability across random-seed configurations.}
To further assess the variability of the numerical results across different random-seed configurations, Tables~\ref{tab:exp_volume} and~\ref{tab:exp_time} report the mean and standard deviation of the certified volumes and computation times, respectively, over the three random-seed configurations described in Appendix~\ref{sec:app_exp}. These results complement the averaged results reported in Table~\ref{tab:exp} of the main text.

\begin{table*}[!htb]
\centering
\caption{Relative certified volumes across three seed configurations.}
\setlength{\tabcolsep}{0.8mm}{
\begin{tabular}{ccccccc}
\hline
\multirow{2.5}{*}{Ex}
& \multirow{2.5}{*}{\parbox{1.8cm}{\centering Robust\\invariant set}}
& \multirow{2.5}{*}{\parbox{1.6cm}{\centering Worst-case\\ RBGF}}
& \multicolumn{3}{c}{Ours}
& \multirow{2.5}{*}{\parbox{1.8cm}{\centering Finite Time\\~\cite{wu2026pac}}} \\
\cmidrule(lr){4-6}
& &
& $\widehat V_{\text{max}}$ & $\widehat V_{0.8}$ & $\widehat V_{0.5}$
& \\
\hline
1 & $0.100\!\pm\!0.000$ & $0.100\!\pm\!0.000$ & $0.856\!\pm\!0.032$ & $0.428\!\pm\!0.012$ & $0.742\!\pm\!0.026$ & $1.000\!\pm\!0.000$ \\
2 & $0.669\!\pm\!0.023$ & $0.786\!\pm\!0.018$ & $0.922\!\pm\!0.022$ & $0.889\!\pm\!0.022$ & $0.922\!\pm\!0.022$ & $0.898\!\pm\!0.017$ \\
3 & $0.519\!\pm\!0.128$ & $0.519\!\pm\!0.128$ & $1.000\!\pm\!0.000$ & $0.904\!\pm\!0.020$ & $0.998\!\pm\!0.000$ & $0.999\!\pm\!0.000$ \\
4 & $0.299\!\pm\!0.001$ & $0.299\!\pm\!0.001$ & $0.637\!\pm\!0.005$ & $0.570\!\pm\!0.004$ & $0.636\!\pm\!0.005$ & $0.694\!\pm\!0.002$ \\
5 & $0.650\!\pm\!0.005$ & $0.650\!\pm\!0.005$ & $0.903\!\pm\!0.002$ & $0.890\!\pm\!0.002$ & $0.903\!\pm\!0.002$ & $0.927\!\pm\!0.001$ \\
6 & $0.517\!\pm\!0.016$ & $0.888\!\pm\!0.001$ & $1.000\!\pm\!0.000$ & $0.982\!\pm\!0.001$ & $1.000\!\pm\!0.000$ & $1.000\!\pm\!0.000$ \\
\ref{ex:rotor3d} & $0.653\!\pm\!0.000$ & $0.790\!\pm\!0.119$ & $0.976\!\pm\!0.000$ & $0.964\!\pm\!0.000$ & $0.976\!\pm\!0.000$ & $1.000\!\pm\!0.000$ \\
\ref{ex:shg4d} & $0.644\!\pm\!0.000$ & $0.648\!\pm\!0.000$ & $0.988\!\pm\!0.006$ & $0.972\!\pm\!0.006$ & $0.988\!\pm\!0.006$ & $1.000\!\pm\!0.000$ \\
\ref{ex:6d} & $0.065\!\pm\!0.000$ & $0.065\!\pm\!0.000$ & $1.000\!\pm\!0.000$ & $0.191\!\pm\!0.000$ & $0.725\!\pm\!0.000$ & $1.000\!\pm\!0.000$ \\
\hline
\end{tabular}}
\label{tab:exp_volume}
\end{table*}

\begin{table*}[!htb]
\centering
\caption{Computation times in seconds across three seed configurations.}
\setlength{\tabcolsep}{2mm}{
\begin{tabular}{ccccc}
\hline
Ex
& Robust invariant set
& Worst-case RBGF
& Ours
& Finite Time~\cite{wu2026pac} \\
\hline
1 & $0.0\pm0.0$ & $1.0\pm0.2$ & $247\pm65$  & $220\pm85$ \\
2 & $1.2\pm0.3$ & $2.3\pm0.2$ & $621\pm133$ & $251\pm89$ \\
3 & $2.4\pm0.3$ & $6.9\pm0.2$ & $918\pm88$  & $468\pm20$ \\
4 & $1.7\pm0.2$ & $3.1\pm0.7$ & $975\pm83$  & $517\pm35$ \\
5 & $0.9\pm0.2$ & $1.2\pm0.1$ & $98\pm17$   & $69\pm4$ \\
6 & $1.6\pm0.0$ & $3.0\pm0.7$ & $489\pm31$  & $505\pm74$ \\
\ref{ex:rotor3d} & $0.5\pm0.1$ & $0.7\pm0.1$ & $23\pm1$ & $25\pm1$ \\
\ref{ex:shg4d} & $0.6\pm0.0$ & $0.7\pm0.0$ & $73\pm4$ & $62\pm3$ \\
\ref{ex:6d} & $0.0\pm0.0$ & $1.2\pm0.1$ & $519\pm52$  & $589\pm101$ \\
\hline
\end{tabular}}
\label{tab:exp_time}
\end{table*}

\end{document}